\documentclass[fleqn, usenatbib]{mnras}

\usepackage{newtxtext,newtxmath}

\usepackage[T1]{fontenc}
\usepackage{ae,aecompl}
\usepackage{gensymb}
\usepackage{acronym}
\acrodef{HST}[\emph{HST}]{\emph{Hubble Space Telescope}}
\acrodef{ICM}{intra-cluster medium}
\acrodef{SF}{star-forming}
\acrodef{ISM}{interstellar medium}
\acrodef{BCG}{brightest cluster galaxy}
\acrodef{SFR}{star formation rate}
\acrodef{sSFR}{specific star formation rate}
\acrodef{AGN}{active galactic nuclei}
\acrodef{DM}{dark matter}
\acrodef{CGM}{circumgalactic medium}
\acrodef{IMF}{initial mass function}
\acrodef{WFC3}{Wide Field Camera 3}
\acrodef{GLASS}{Grism Lens-Amplified Survey from Space}
\acrodef{ESO-NTT}{European Southern Observatory New Technology Telescope}
\acrodef{SN}{supernova}
\acrodef{MUSE}{Multi Unit Spectroscopic Explorer}
\acrodef{MOSFIRE}{Multi-Object Spectrometer For Infra-Red Exploration}
\acrodef{SSP}{simple stellar population}
\acrodef{RPS}{Ram pressure stripping}
\acrodef{FOV}{field-of-view}

\usepackage{graphicx}   
\usepackage{amsmath}    
\usepackage{siunitx}    
    \DeclareSIUnit \parsec {pc}
    \DeclareSIUnit \year {yr}
    \DeclareSIUnit \AU {AU}
    \DeclareSIUnit \erg {erg}
    \DeclareSIUnit \solarmass {M_{\odot}}
    \DeclareSIUnit \angstrom {\text {Å}}
\usepackage{dirtytalk} 
\usepackage{url}
\usepackage{enumitem}
\usepackage{subcaption}

\title[Anisotropic quenching in massive clusters]{Evidence that pre-processing in filaments drives the anisotropic quenching of satellite galaxies in massive clusters}

\author[H.M.O. Stephenson et al.]{H. M. O. Stephenson$^{1}$\thanks{E-mail: h.stephenson@lancaster.ac.uk (HMOS)},
J. P. Stott$^{1}$\thanks{E-mail: j.p.stott@lancaster.ac.uk (JPS)},
J. Butler$^{2}$,
M. Webster$^{1}$,
and J. Head$^{1}$
\\
$^{1}$Department of Physics, Lancaster University, Lancaster LA1 4YB, UK\\
$^{2}$School of Physics and Astronomy, University of Nottingham, University Park, Nottingham NG7 2RD, UK\\
}

\date{Accepted XXX. Received YYY; in original form ZZZ}

\pubyear{2024}
\begin{document}
\label{firstpage}
\setlength{\abovedisplayskip}{0pt}
\setlength{\belowdisplayskip}{0pt}
\setlength{\parskip}{0pt}
\pagerange{\pageref{firstpage}--\pageref{lastpage}}
\maketitle

\begin{abstract}

We use a sample of 11 $z\approx0.2-0.5$ ($z_{\text{med.}} = 0.36$) galaxy clusters from the Cluster Lensing And Supernovae survey with Hubble (CLASH) to analyse the angular dependence of satellite galaxy colour $(B-R)$ and passive galaxy fraction ($f_{\text{pass.}}$) with respect to the major axis of the brightest cluster galaxy (BCG). This phenomenon has been dubbed as \say{anisotropic quenching}, \say{angular conformity} or \say{angular segregation}, and it describes how satellite galaxies along the major axis of the BCG are more likely to be quenched than those along the minor axis. A highly significant anisotropic quenching signal is found for satellites, with a peak in $(B-R)$ and $f_{\text{pass.}}$ along the major axis. We are the first to measure anisotropic quenching out to cluster-centric radii of $3R_{200}$ ($R_{200\text{, med.}} \approx 933$ \si{\kilo\parsec}). We find that the signal is significant out to at least $2.5R_{200}$, and the amplitude of the signal peaks at $\approx1.25R_{200}$. This is the first time a radial peak of the anisotropic quenching signal has been measured directly. We suggest that this peak could be caused by a build-up of backsplash galaxies at this radius. Finally, we find that $f_{\text{pass.}}$ is significantly higher along the major axis for fixed values of local surface density. The density drops less rapidly along the major axis and so satellites spend more time being pre-processed here compared to the minor axis. We therefore conclude that pre-processing in large-scale structure, and not active galactic nuclei outflows (AGN), is the cause of the anisotropic quenching signal in massive galaxy clusters, however this may not be the cause in lower mass halos.


\end{abstract}




\section{Introduction}
\label{sec::intro}

Galaxy clusters are the densest and most massive virialized structures in the Universe, often hosting thousands of galaxies. They form from the collapse of large, gravitationally-bound overdensities in the initial density field of the Universe and follow a hierarchical sequence of dark matter halo mergers \citep{Kravtsov2012}, resulting in masses of $M_{\text{h}} \gtrsim 10^{14}$ \si{\solarmass} \citep{Overzier2016}. Permeating in between cluster galaxies is hot, dense gas known as the \ac{ICM}, which can be used to parametrize clusters since the X-ray luminosity and temperature of the gas scales with the overall mass of the system. These dense environments, combined with the hot \ac{ICM}, lead to the quenching of star formation within their galaxies, resulting in cluster galaxies being typically more red \citep{Butcher1984,DeLucia2006}, less \ac{SF} \citep{Chartab2020} and more elliptical \citep{Dressler1980,Postman2005} than those found in the field.

Many different quenching mechanisms can lead to the suppression of star formation. \citet{Peng2010b} found that the effects of a galaxy's stellar mass and its surrounding environment can be cleanly separated into different mechanisms known as mass quenching and environmental quenching respectively. In galaxy clusters, it is environmental quenching mechanisms that dominate the suppression of star formation in satellite galaxies as they orbit within the cluster potential. This is due to the density of the \ac{ICM}, the proximity of galaxies to other satellites and the massive dark matter halo. \ac{RPS} is an environmental quenching mechanism which occurs when a galaxy falls into a cluster and its cold \ac{ISM} is shocked and removed by the much hotter \ac{ICM} \citep{Gunn1972,Boselli2014a}. This strips away the fuel for star formation, leaving the galaxy to quench on a relatively quick timescale of $\sim1-2$ \si{\giga\year} \citep{Roberts2019,Akins2021}. \ac{RPS} may also cause significantly higher \ac{SFR} at the shock front for a brief period which causes the galaxy to burn through fuel quickly, again leading to quenching. Satellite galaxies can also have their gas stripped away as a result of tidal interactions with other galaxies \citep{Richstone1975,Richstone1976}. This will happen if a galaxy experiences a close encounter with a more massive galaxy, such as the central \ac{BCG}. Similarly, a satellite is likely to experience several encounters with other satellite galaxies due to their high number density, leading to the repeated loss of cold gas in a process that is more commonly known as galaxy harassment \citep{Moore1996}. These interactions can also funnel gas towards the centre of satellites which results in a burst of star formation that quickly spends all the cold gas available leading to a period of quiescence \citep{Barnes1996,Barnes2004,Hopkins2013a}. These tidal effects are not just limited to direct interactions between the galaxies themselves, but could also be the result of the potential of the cluster disturbing the thin disk of gas and triggering bursts of star formation \citep{Byrd1990,Moore1999}. Mergers of galaxies can also lead to quenching of star formation \citep{Mihos1994a,Hopkins2008,Poggianti2017}, but this is relatively rare in clusters due to the high-velocity dispersions of $\sigma \sim 1000$ \si{\kilo\meter\per\second} \citep{Struble1999}. 

Satellite galaxies replenish cold gas that is used up in the process of forming stars by accreting it from reservoirs of gas in the galactic halo (see \citealp{Putman2012} for a review). If this supply is cut off from the galaxy, either via heating from the \ac{ICM} (e.g. \citealp{Dekel2006}) or removed from the galaxy's dark matter halo (e.g. \citealp{Vaughan2020}), then any gas in the \ac{ISM} is used up over time with no replenishment and the galaxy will become quiescent on timescales of up to 4 \si{\giga\year} \citep{Roberts2019} in a process known as `starvation' or `strangulation' \citep{Larson1980}.

In clusters, the distribution of satellite galaxies around the central cluster galaxy (typically the \ac{BCG}) is anisotropic such that there exists a relative overdensity of galaxies along a BCG's major axis compared to its minor axis. This is known as BCG-cluster alignment \citep{Sastry1968,Brainerd2005}. This signal has been observed in both observations (e.g. \citealp{Carter1980,Wang2008,Huang2016}) and simulations (e.g. \citealp{Kang2007,Ragone-Figueroa2020,Gu2022}), and is thought to arise from a combination of the preferential infall of satellite galaxies along cosmic filaments that align with the \ac{BCG} (e.g. \citealp{Libeskind2011,Libeskind2013,Welker2018,Smith2023}), primordial alignment during the formation of the \ac{BCG} (e.g. \citealp{West1994}) and satellite galaxies gradually aligning with the local tidal field as a result of gravitational torques (e.g. \citealp{Catelan1996}). These distributions could be linked to the anisotropic shape of the overall \ac{DM} halo which tends to be triaxial, as seen in both cosmological simulations \citep{Frenk1988,Vega-Ferrero2017} and observations \citep{Sereno2013,Gonzalez2021}. \ac{DM} haloes have been found to become more ellipsoidal with increasing cluster mass as well \citep{Despali2017,Okabe2019}. The mass of a cluster may also directly influence the satellite distribution around the \ac{BCG} \citep{Paz2006,Paz2011}. See \citet{Kirk2015} for a more detailed review of these galaxy alignments in observations, and \citet{Kiessling2015} for a review in simulations.

Within the aforementioned anisotropic distribution of satellites is a more clear alignment of red satellites along the major axis of the central galaxy. \citet{Yang2006} find an excess of satellites along the major axes of their Sloan Digital Sky Survey (SDSS; \citealp{York2000}) galaxy groups. They find that the signal is strongest when they only consider those groups with red central galaxies and red satellites, a result that was repeated soon after by \citet{Azzaro2007} for isolated hosts using SDSS DR4 \citep{AdelmanMccarthy2006}. \citet{Huang2016} used the redMaPPer cluster catalogue \citep{Rykoff2014} to analyse the satellite distribution in galaxy clusters, and they too found that red satellites are more likely to reside along the major axis of the central cluster galaxy, concluding that it is colour that is the strongest predictor to what they call \say{angular segregation}. This stronger alignment of red satellites has also been seen in simulations (e.g. \citealp{Dong2014}). This preferential alignment of red satellites could suggest that there are quenching mechanisms that seem to favour the major axis of the central galaxy. This has recently been dubbed \say{anisotropic quenching} or \say{angular conformity} \citep{Martin-Navarro2021}.

\citet{Martin-Navarro2021} analysed 124,000 $z\approx0.08$ satellite galaxies from SDSS DR10 \citep{Ahn2014} using galaxy group/cluster catalogues built by \citet{Tempel2014}. They found that satellites positioned along the minor axis of the central galaxy were less quenched compared to those along the major axis, measured using a change in the quiescent fraction of galaxies. The authors attributed this anisotropic quenching to minor axis outflows from the \ac{AGN} activity of the central galaxy, creating bubbles of low-density gas in the \ac{CGM}. The efficiency of \ac{RPS} within these bubbles is greatly reduced which leads to a lower fraction of quenched galaxies along this axis. \citet{Martin-Navarro2021} solidify these observations by reproducing the signal in the IllustrisTNG cosmological hydrodynamical simulations \citep{Nelson2019}, and by also observing a stronger signal in SDSS for satellites orbiting central galaxies with more massive supermassive black holes. These results were later supported by \citet{Zhang2022} who found a drop in the emission line flux of [OIII] and H$\beta$ from the major to the minor axis which indicates a drop in the density of the \ac{CGM} which they link to \ac{AGN} activity (though their results are statistically marginal and are only qualitatively consistent with those of \citealp{Martin-Navarro2021}).

\citet{Stott2022} found the same result as \citet{Martin-Navarro2021} but extended the observation out to $z\sim0.5$ using the Cluster Lensing And Supernova Survey with Hubble (CLASH; \citealp{Postman2012}). He found a significant anisotropic quenching signal for his sample of $z=0.391-0.545$ clusters but was limited by the narrow field of view of the \ac{HST}'s \ac{WFC3} \citep{MacKenty2008,MacKenty2010} for his lower redshift sample. He found an anisotropic quenching signal in both average galaxy colour and quenching fraction, noting that galaxies bluer than the red sequence were less common in a region $\pm45\degree$ from the major axis of the \ac{BCG}. \citet{Stott2022} offers an alternative suggestion to \citet{Martin-Navarro2021} for the observed signal: since the shape and major axis of both the cluster and \ac{BCG} typically align with each other (e.g. \citealp{Binggeli1982,West1994}), the \ac{ICM} density at a fixed radius from the \ac{BCG} would be higher along the major axis. Conversely, along the minor axis, the lower density leads to a reduction in the efficiency of environmental quenching mechanisms in the minor axis plane of the \ac{BCG}. Therefore, an anisotropic quenching signal may arise from the fact that these ellipsoidal effects are studied within a circular cluster-centric radius that does not take into account the non-circular distribution of the \ac{ICM}. However, he notes that the CLASH clusters are not highly elliptical (average ellipticity $\overline{\epsilon}=0.19$; \citealp{Postman2012}). On a related point, he suggests it may be due to the increased galaxy density on the major axis that increases galaxy interactions.

Going out to $z\sim1$, \citet{Ando2023} detected anisotropic quenching in $z=0.25-1$ CAMIRA \citep{Oguri2014} clusters in the Hyper Suprime-Cam \citep{Miyazaki2012} Subaru Strategic Program (HSC-SSP; \citealp{Aihara2018,Aihara2022}) S20A data \citep{Oguri2018}, but find no evidence of a signal at $z>1$. They see quiescent fractions along the major axis are consistently higher within $R_{\text{200}}$, but find the fractions are similar along both axes at larger radii, which implies physical mechanisms that only operate within the \ac{DM} halo of the cluster are responsible for anisotropic quenching. They go further by ruling out the cause of anisotropic signal being differences in the local density of satellite galaxies along both axes because the quiescent fraction remains higher along the major axis when this parameter is fixed. Additionally, they find that the excess in the quiescent fraction in the major axis compared to the minor axis is independent of stellar mass. Given low mass galaxies are more susceptible to \ac{RPS} \citep{Gunn1972,Steyrleithner2020}, this suggests a different quenching process may be responsible instead, contradicting the theory proposed by \citet{Martin-Navarro2021}. However, \citet{Ando2023} note that their stellar mass cut may be too high to definitively conclude this.

In this paper, we aim to further probe anisotropic quenching in galaxy clusters by analysing Subaru observations of the CLASH clusters at $z=0.206-0.494$ \citep{Postman2012,Umetsu2014}. We measure both the colour and the quiescent fraction of satellite galaxies as a function of the orientation angle from the \ac{BCG} major axis. One of the main aims of this work is to extend the analysis of anisotropic quenching out to larger cluster-centric radii than was possible for \citet{Stott2022} - who also analysed the CLASH clusters - by utilising the much larger \ac{FOV} of the Subaru Telescope compared to \ac{HST}-\ac{WFC3}. This allows us to probe out to 3$R_{200}$ ($\approx 2200-3200$ \si{\kilo\parsec}) for the clusters in our sample. We also perform a test to see if any detected signal is dependent on the local density of satellite galaxies along both axes, by measuring the overall number density of satellites as well as the average number density within the average area to the 4th- and 5th- nearest neighbours of each satellite.

This paper is arranged as follows. In Section \ref{sec::data}, the CLASH survey is described and the sample used for this work is explained in detail. The results are outlined in Section \ref{sec::results}. A discussion of the results can be found in Section \ref{sec::discussion}. Our conclusions are summarised in Section \ref{sec::conclusions}.

A standard $\Lambda$CDM cosmology model is assumed with values $\Omega_{\Lambda} = 0.7$, $\Omega_{m} = 0.3$, $H_{0} = 70$ \si{\kilo\meter\per\second\per\mega\parsec}. Any magnitudes stated are presented using the AB system. All results and models in this work assume a \citet{Chabrier2003} \ac{IMF} throughout. The clusters will be referred to using abbreviations of their full name: Abell clusters will still be referred to as Abellxxxx (e.g. Abell611); the MAssive Cluster Survey\footnote{https://home.ifa.hawaii.edu/users/ebeling/clusters/MACS.html} (MACS) clusters (see \citealp{Ebeling2001}) will be referred to as MACSxxxx (e.g. MACS1206.2-0847 will be MACS1206); and RXJ2129+0005, RXJ1532.9+3021 \& RXJ1347-1145 will be abbreviated to RXJxxxx.


\section{Sample and Data}
\label{sec::data}

\begin{table*}
    
    \footnotesize
    \centering
    \caption{Our sample of CLASH clusters. The right ascension (R.A.) and declination (Dec.) of the BCG are derived from the Subaru observations performed by \citet{Umetsu2014}. The $M_{R_{200}}$, Scale Radius ($r_{s}$) and $R_{200}$ Concentration Parameter ($c_{R_{200}}$) data is taken from Table 7 in \citet{Merten2015}. The $R_{200}$ data is calculated by $R_{200} = c_{R_{200}} \cdot r_{s}$.}
    \label{tab::clashinfo}
    \begin{tabular}{c|c|c|c|c|c|c|c|c}
    \hline
    Cluster & $z_{\text{s}}$ & BCG R.A. & BCG Dec. & $M_{R_{200}}$ ($10^{14}$ \si{\solarmass}) & $r_{s}$ (\si{\kilo\parsec}) & $c_{R_{200}}$ & $R_{200}$ (\si{\kilo\parsec}) & Filters \\ \hline
    Abell209    & 0.206 & 01:31:52.56 & -13:36:40.32 & $6.65\pm0.49$   & 322 $\pm$ 49  & 3.3 $\pm$ 0.9    & 1063 $\pm$ 332  & $B_{J}, V_{J}, R_{C}, i^{\prime}, z^{\prime}$       \\
    RXJ2129     & 0.234 & 21:29:39.96 & 00:05:20.70  & $4.27\pm0.42$   & 210 $\pm$ 35  & 4.3 $\pm$ 1.4    & 903 $\pm$ 330   & $B_{J}, V_{J}, R_{C}, i^{\prime}, z^{\prime}$       \\
    Abell611    & 0.288 & 08:00:57.02 & 36:03:28.30  & $5.95\pm0.35$   & 287 $\pm$ 42  & 3.4 $\pm$ 0.9    & 976 $\pm$ 295   & $B_{J}, V_{J}, R_{C}, I_{C}, i^{\prime}, z^{\prime}$\\
    MACS2137    & 0.313 & 21:40:15.10 & -23:39:38.63 & $7.28\pm0.42$   & 336 $\pm$ 35  & 3.1 $\pm$ 0.6    & 1042 $\pm$ 229  & $B_{J}, V_{J}, R_{C}, I_{C}, z^{\prime}$         \\
    RXJ1532     & 0.345 & 15:32:53.76 & 30:21:00.43  & $3.71\pm0.56$   & 273 $\pm$ 70  & 3.0 $\pm$ 1.4    & 819 $\pm$ 436   & $B_{J}, V_{J}, R_{C}, I_{C}, z^{\prime}$         \\
    MACS1931    & 0.352 & 19:31:49.61 & -26:34:34.43 & $4.83\pm0.35$   & 287 $\pm$ 49  & 3.2 $\pm$ 0.9    & 918 $\pm$ 302   & $B_{J}, V_{J}, R_{C}, I_{C}, z^{\prime}$         \\
    MACS1115    & 0.352 & 11:15:51.89 & 01:29:57.69  & $6.30\pm0.63$   & 434 $\pm$ 77  & 2.3 $\pm$ 0.7    & 998 $\pm$ 352   & $B_{J}, V_{J}, R_{C}, I_{C}, z^{\prime}$         \\
    MACS1720    & 0.391 & 17:20:16.92 & 35:36:26.50  & $5.25\pm0.56$   & 217 $\pm$ 42  & 4.3 $\pm$ 1.4    & 933 $\pm$ 353   & $B_{J}, V_{J}, R_{C}, I_{C}, z^{\prime}$         \\
    MACS1206    & 0.440 & 12:06:12.10 & -08:48:05.24 & $6.02\pm0.77$   & 217 $\pm$ 42  & 4.3 $\pm$ 1.5    & 933 $\pm$ 372   & $B_{J}, V_{J}, R_{C}, I_{C}, z^{\prime}$         \\
    MACS0329    & 0.450 & 03:29:41.55 & -02:11:44.13 & $5.11\pm0.70$   & 231 $\pm$ 56  & 3.8 $\pm$ 1.6    & 878 $\pm$ 427   & $B_{J}, V_{J}, R_{C}, I_{C}, z^{\prime}$         \\
    MACS1311    & 0.494 & 13:11:01.78 & -03:10:37.49 & $3.22\pm0.21$   & 168 $\pm$ 21  & 4.4 $\pm$ 1.0    & 739 $\pm$ 192   & $B_{842}, V_{843}, R_{C}, z^{\prime}_{\text{IMACS}}$ $^{a}$   \\ \hline
    \end{tabular}
\begin{flushleft}
$^{a}$ - For the equivalent $B$ and $V$ bands in MACS1311 observations, \citet{Umetsu2014} make use of data from the Wide-Field Imager on the European Southern Observatory’s MPG/ESO telescope \citep{Baade1999} ($B_{842}$ and $V_{843}$ respectively). For the equivalent $z^{\prime}$ band, observations from the Magellan-Baade telescope's Inamori–Magellan Areal Camera and Spectrograph (IMACS; \citealp{Dressler2011}) are used ($z^{\prime}_{\text{IMACS}}$).
\end{flushleft}
\end{table*}











Our work focuses on observations of galaxy clusters from CLASH (\emph{HST} Cycle 18 Multi-Cycle Treasury Program GO-12065; PI: M. Postman; see \citealp{Postman2012}). CLASH imaged 25 massive galaxy clusters at intermediate redshifts ($z=0.187-0.890$) from 524 \ac{HST} orbits in order to establish their mass and DM concentrations via gravitational lensing \citep{Postman2012,Umetsu2012}. Of these 25 clusters, 5 were selected based on their gravitational lensing properties (to increase the probability of discovering $z>7$ galaxies) and 20 were X-ray selected to limit the effect of lensing biases in their overall sample \citep{Postman2012}.

The data for our work are from observations using the Subaru Prime Focus Camera (Suprime-Cam) mounted on the wide-field prime focus of the Subaru Telescope. Suprime-Cam is an 80-mega pixel ($10240 \times 8192$) mosaic CCD camera and is built for extremely wide-field images. The camera covers a \ac{FOV} of $34' \times 27'$ with a resolution of 0''.202 per pixel \citep{Miyazaki2002}. The Suprime-Cam data was primarily analysed by \cite{Umetsu2014} for their investigation into the joint shear-and-magnification weak-lensing of 20 CLASH clusters between $z=0.187-0.690$, and is publicly available on the CLASH website\footnote{https://archive.stsci.edu/prepds/clash/}. Each of the clusters was observed in at least three optical bands and up to six total bands for some clusters, with exposure times of between $1000-10000$ \si{\second} per passband. The six broad-band filters used for these observations were $B_{\text{J}}$ and $V_{\text{J}}$ from the Johnson-Morgan system \citep{Johnson1953}, $R_{\text{C}}$ and $I_{\text{C}}$ from the Cousins system \citep{Cousins1978} and the $i^{\prime}$ and $z^{\prime}$ filters from the SDSS system \citep{Fukugita1996}. These six filters cover a total spectral range of $\sim3600-10700$ \si{\angstrom} (with the exception of MACS1311 which only had $R_{C}$ data available from Subaru; see Table \ref{tab::clashinfo}). \citet{Umetsu2014} present most of their results in $R_{\text{C}}$, and describe the typical limiting magnitudes as $\sim26-26.5$ mag in this band for a $3\sigma$ detection. Further independent analysis of these clusters as part of the `\emph{Weighing the Giants}' project, for which a substantial fraction of this data was taken, can be found in the series of papers cataloguing their results \citep{VonDerLinden2014,Kelly2014,Applegate2014,Mantz2015,Mantz2016}.

We used the CLASH catalogue generated by \citet{Molino2017} to acquire the position angles of the \ac{BCG} in order to analyse the angular dependence of galaxy colour and passive galaxy fraction ($f_{\text{pass.}}$). The data from \citet{Molino2017} is also available on the CLASH website. We determine the \ac{BCG} from visual inspection of the Subaru images, selecting the brightest galaxy at the same redshift of the cluster, typically with a cD halo, and close to the X-ray centroid of the observations (excluding clusters with ambiguous \ac{BCG}s, see below). The photo-$z$ estimates for the Subaru data are derived by \citet{Umetsu2014} who ran the Bayesian photometric redshift estimation code \citep{Benitez2000} and corrected for Galactic extinction following the dust maps of \citet{Schlegel1998}.

Of the 25 original CLASH clusters, we determined that 11 were suitable for analysis in this work. Primarily, we wanted to focus on curating a sample which uses a consistent colour index that would correlate well with \ac{SFR}. This colour index should therefore have magnitudes from filters that are on either side of the rest-frame 4000 \si{\angstrom} break i.e. an approximation of rest-frame $U-V$. The 4000 \si{\angstrom} break is often used as a dividing line between young, \ac{SF} galaxy populations and older, quiescent galaxies as a result of an accumulation of absorption lines from metals and the Balmer series at this wavelength. As stellar populations age and cool, the break becomes larger as they get more opaque, making it easier to determine if a galaxy is actively forming new stars (e.g. \citealp{Dressler1987,Kauffmann2003a,Kriek2011}). In our case, the best-observed colour index is $B_{J} - R_{C}$ (hereafter $B-R$) and, as a result of the evolving position of the break moving beyond the wavelength range of these filters, 7 clusters at $z\gtrsim0.5$ are excluded. Abell1423 and RXJ2248 were removed because there is no available Subaru data. Abell383 was removed because the maximum fraction of $R_{200}$ that could be probed with Subaru ($\approx 2.75R_{200}$) was deemed to be too restrictive compared to the higher redshift clusters ($\geq3R_{200}$). Following visual inspection, the BCGs in Abell2261 and MACS0416 had multiple identified cores - possibly as a result of a merger - which made it difficult to determine the position angle so they were excluded. Similarly, RXJ1347 has multiple possible BCGs so this was also excluded. Finally, MACS0429 was removed as no $B_{J}$ data was taken in the Subaru observations. The details of our final cluster sample can be found in Table \ref{tab::clashinfo}. The median $\text{M}_{R_{200}}$ of our final cluster sample, using the cluster masses determined by \citet{Merten2015}, is $M_{R_{200}, \text{med.}} = (5.25 \pm 1.20)\times10^{14}$ \si{\solarmass} with a median $R_{200\text{, med.}} = 933 \pm 90$ \si{\kilo\parsec}.

To determine cluster membership, we applied an evolving photometric redshift restriction on galaxies in the Subaru catalogue. This designation was determined following analysis of the scatter of the photometric redshifts in \citet{Umetsu2014} compared to spectroscopically confirmed redshifts of the same sources. Spectroscopic redshifts are available for cluster members in Abell209 \citep{Annunziatella2016}, MACS1206 \citep{Biviano2013} and MACS2129 \citep{Monna2017} following observations by the CLASH-VLT program (ESO VLT Large programme 186.A-0798; PI: P. Rosati; see \citealp{Rosati2014}). For cluster members of MACS0717, MACS0744, MACS1149, MACS1423 and RXJ1347, spectroscopic redshifts are available from \ac{GLASS}\footnote{https://archive.stsci.edu/prepds/glass/} (\ac{HST} Cycle 21 Large Program 13459; PI: T. Treu; see \citealp{Schmidt2014,Treu2015}). Additional spectroscopic redshifts are acquired for known sources in two of the clusters in our final sample from the literature: for Abell209, spectroscopic redshifts are acquired from \ac{ESO-NTT} observations by \citet{Mercurio2003,Mercurio2008}; for MACS1149, spectroscopic follow-ups were taken to study the \ac{SN} \say{Refsdal} using both \ac{MUSE} (ESO prog.ID 294.A-5032; PI: C. Grillo; see \citealp{Grillo2016}) and Multi-Object Spectrometer For Infra-Red Exploration (MOSFIRE; \citealp{Brammer2016}) which generated additional redshifts (see \citealp{Treu2016} for a summary of the MACS1149 catalogues). Sources with spectroscopic redshifts were then matched to the photo-$z$ catalogues of \citet{Umetsu2014}. The standard deviation of the photo-$z$ at each respective cluster's redshift was measured and we fit a linear relationship in the form of
\begin{equation*}
    \log_{10}(2\cdot\sigma_{\text{ph}z}) = (3.26\pm1.05)\cdot\log_{10}(1+z_{\text{cluster}}) - (1.56\pm0.19),
\end{equation*}

where $z_{\text{cluster}}$ is the spectroscopic redshift of the cluster and $\sigma_{\text{ph}z}$ is the standard deviation of the photo-$z$ measurements at $z_{\text{cluster}}$. We used $2\cdot\sigma_{\text{ph}z}$ to be confident we are taking into account the uncertainty on the photo-$z$ measurements. Therefore, based on this fit, we designated any galaxy to be a cluster member if they were within $\Delta z = 0.03 \cdot (1+z_{\text{cluster}})^{3.26}$.

We also chose to apply an $R$-band completeness limit such that only cluster members with an $R$-band absolute magnitude $\lesssim -16.8$ mag were included, which corresponds to a stellar mass of $\log_{10}(M_{*}/\text{M}_{\odot}) \gtrsim 9.3$. This limit was determined by taking the 95 per cent completeness limit in apparent $R$-band magnitude of the highest redshift cluster in our final sample, MACS1311 ($z = 0.494$), which came to be $\approx 25.6$ mag. Analysis was also done on a sample with a limit corresponding to $\log_{10}(M_{*}/\text{M}_{\odot}) > 10$, or $\lesssim -18.6$ mag in absolute $R$-band at $z=0.494$, to test the effects of stellar mass cuts on the anisotropic quenching signal.



\section{Results}
\label{sec::results}

The principal aim of this study was to analyse the effect that a satellite galaxy's position angle from the \ac{BCG} major axis had on its star formation. We do this in two ways: looking at the angular distribution of the corrected $(B-R)_{\text{corr.}}$ colour of satellites (Section \ref{subsec::anisocolour}), and the quiescent fraction of satellites in the same angular bins (Section \ref{subsec::anisopassive}).

\subsection{Colour Relationship}
\label{subsec::anisocolour}

To probe the colour distribution of \ac{BCG} satellites, we used the colour index $(B-R)_{\text{corr.}}$ (see Section \ref{sec::data}). These clusters cover a redshift range $z=0.206-0.494$, meaning the rest-frame magnitudes are affected by $k$-corrections and possible evolution of the galaxy populations between the upper and lower redshift limits ($\approx2.2$ \si{\giga\year}). To account for this evolution, we apply a $k$- and evolution-correction to shift the clusters to the median redshift of the sample ($z_{\text{median}} \approx 0.362$) using a \ac{SSP} model from \citet{Bruzual2003} with a formation redshift of $z_{f} = 2$, \citet{Chabrier2003} \ac{IMF} and solar metallicity ($Z_{\odot} = 8.69$; \citealp{Asplund2009}). This model is appropriate for quiescent cluster galaxies but not particularly for \ac{SF} galaxies, however the small redshift range means the corrections are suitable for the majority of cases (see also \citealp{Stott2022}).

\begin{figure*}
\centering

\begin{subfigure}{0.45\linewidth}
\includegraphics[width=\linewidth, trim=30 0 50 0]{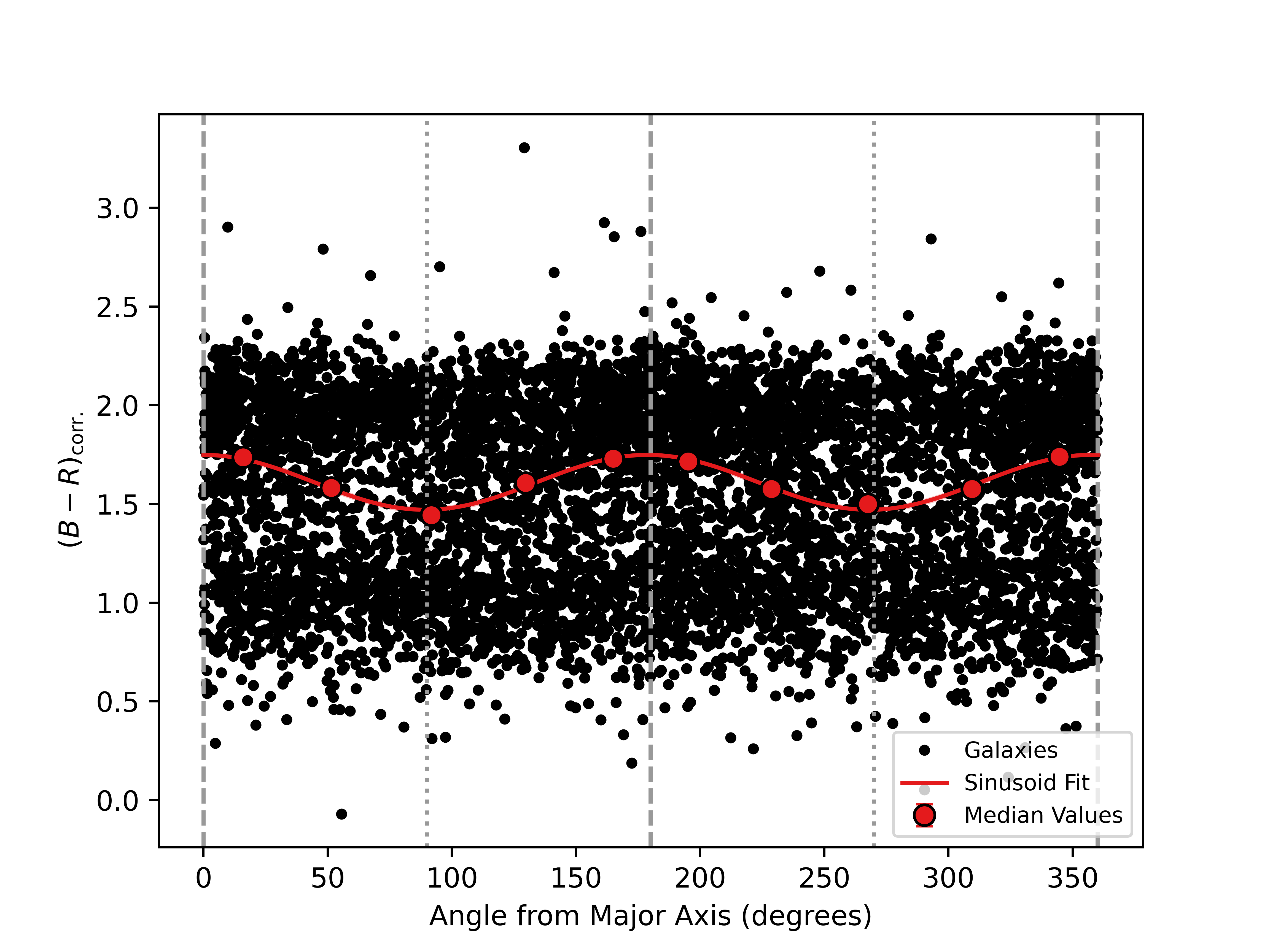}
\caption{ }
\label{subfig::1_5R200_radial_aniso_fullpop_low}
\end{subfigure}
\hfill
\begin{subfigure}{0.45\linewidth}
\includegraphics[width=\linewidth, trim=50 0 30 0]{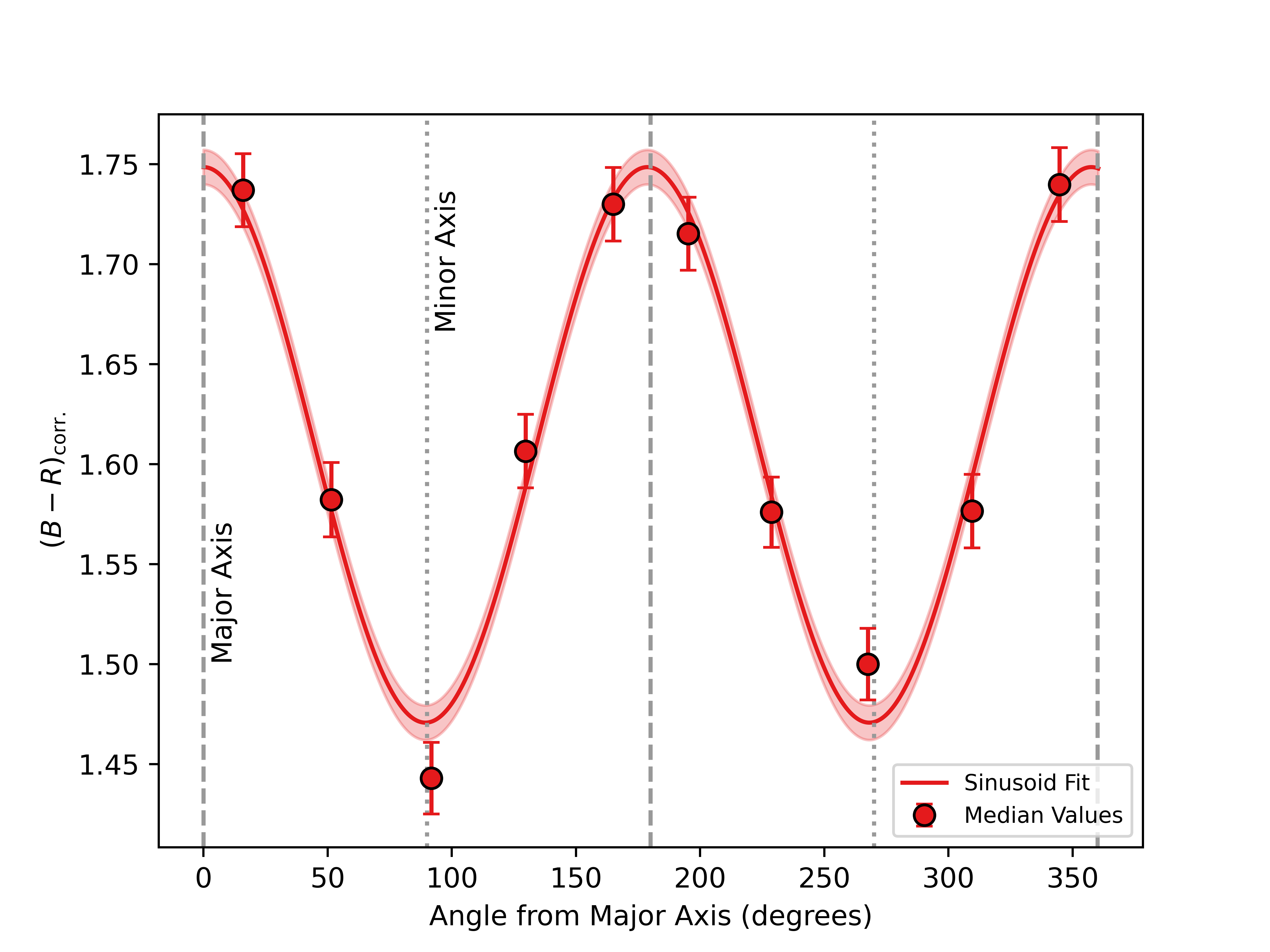}
\caption{ }
\label{subfig::1_5R200_radial_aniso_low}
\end{subfigure}
\begin{subfigure}{0.45\linewidth}
\includegraphics[width=\linewidth, trim=30 0 50 0]{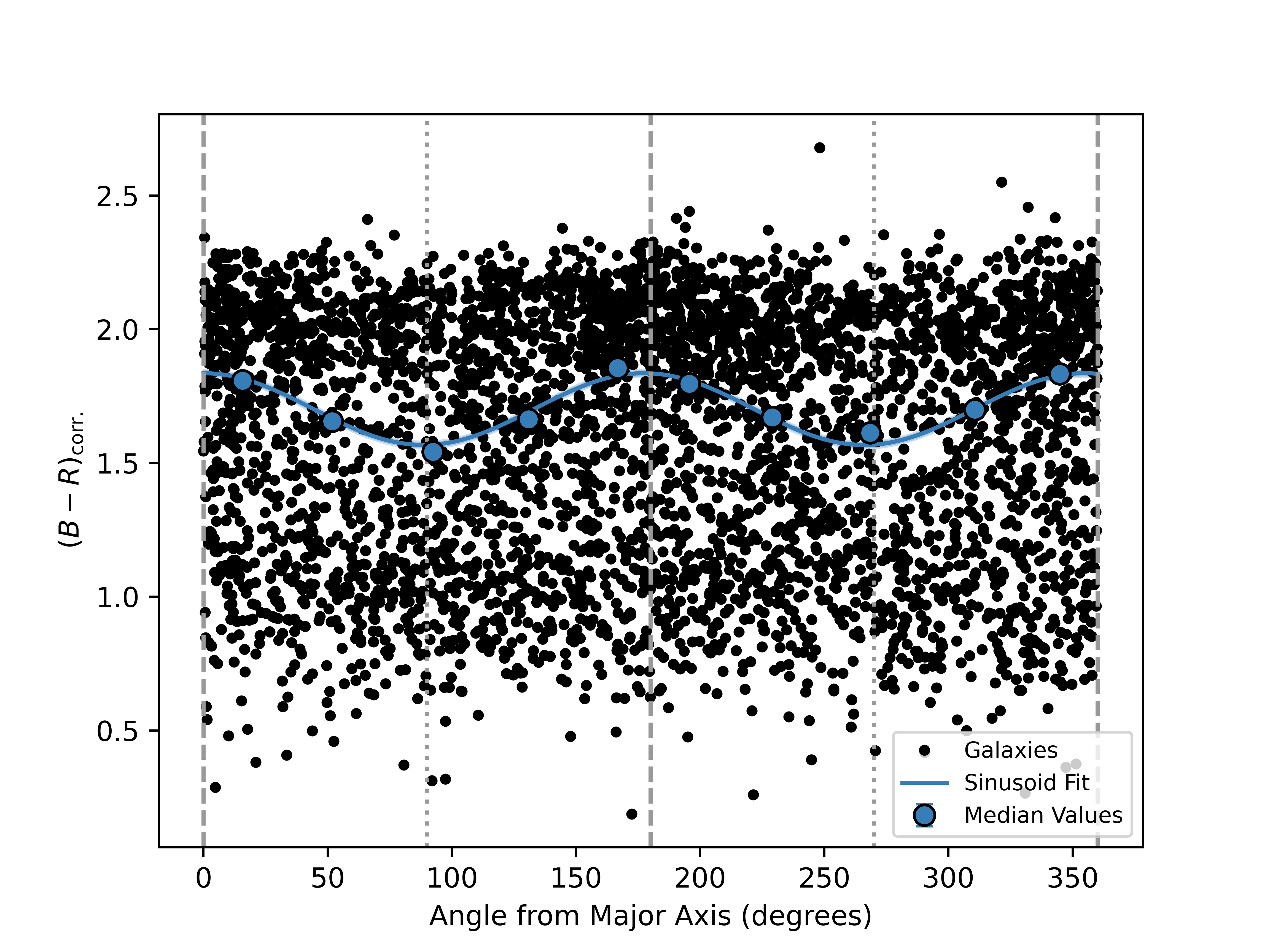}
\caption{ }
\label{subfig::1_5R200_radial_aniso_fullpop_high}
\end{subfigure}
\hfill
\begin{subfigure}{0.45\linewidth}
\includegraphics[width=\linewidth, trim=50 0 30 0]{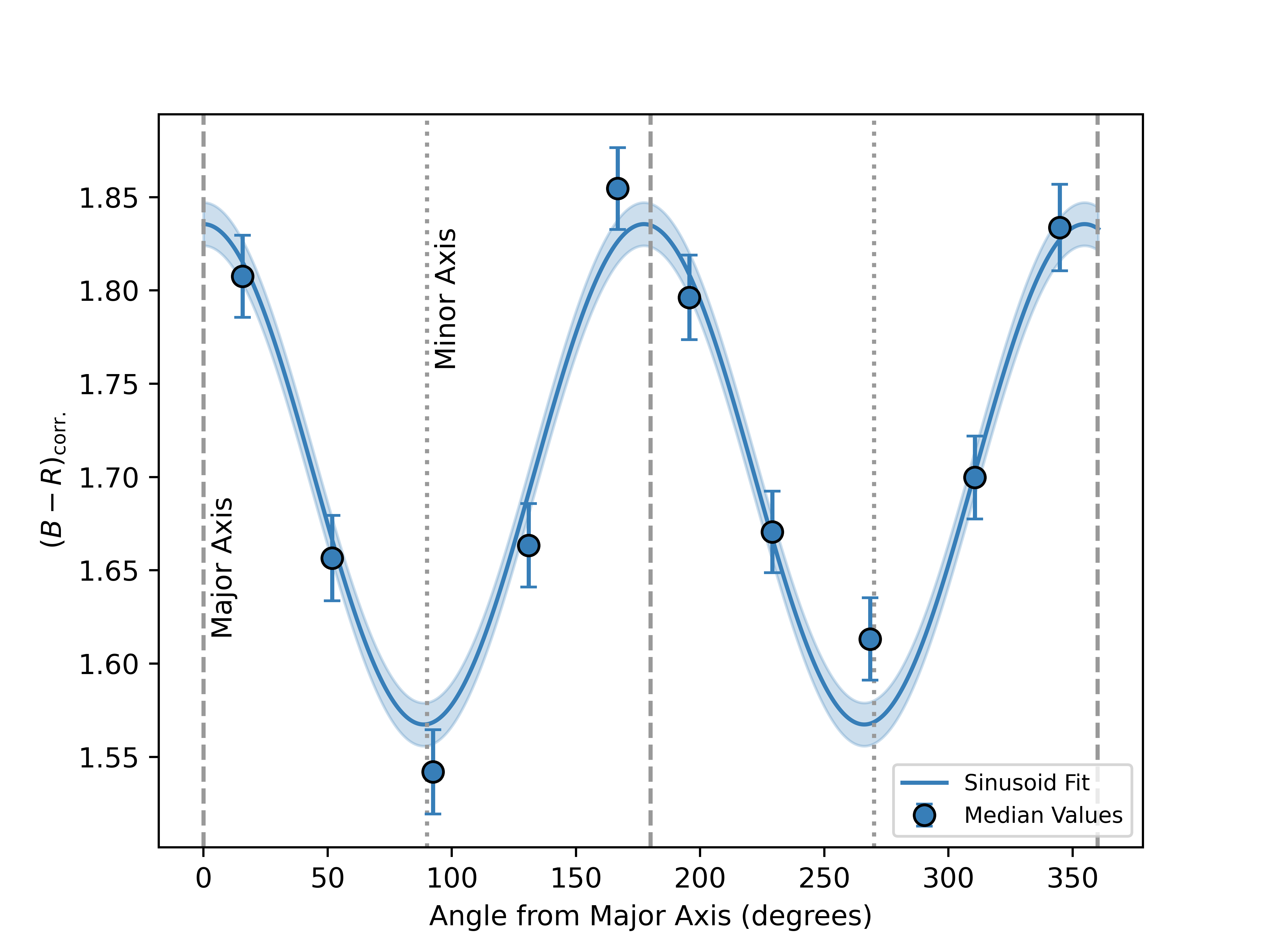}
\caption{ }
\label{subfig::1_5R200_radial_aniso_high}
\end{subfigure}

\caption{Anisotropic quenching signal in colour-angle space for satellite galaxies within $1.5R_{200}$. The left panels show the sinusoidal signal overlayed on the whole population of individual satellites (black points) in this plane, and the right panels just show the binned medians and the sinusoid fits. The red points and fit in Figure \ref{subfig::1_5R200_radial_aniso_low} are the anisotropic signal using an absolute $R$-band completeness limit of $-16.8$ mag, and the blue points and fit in Figure \ref{subfig::1_5R200_radial_aniso_high} use a more restrictive limit of $-18.6$ mag. The $(B-R)_{\text{corr.}}$ errors for both red and blue points represents the standard error within that bin. The shaded regions represent the $1\sigma$ error of the fits. The grey dashed lines indicate the angle at which satellites are along the major axis of the \ac{BCG}, and the grey dotted lines indicate the angle along the minor axis of the \ac{BCG}.}
\label{fig::colourangle_1_5R200}
\end{figure*}

We plotted $(B-R)_{\text{corr.}}$ against angle from the \ac{BCG} major axis in degrees and then binned the median colour in bins of $\approx35-40\degree$. This range in the bin widths is a result of keeping the number of galaxies in each bin approximately constant. We did this using both a $-16.8$ mag completeness limit in absolute $R$-band and a sample with a more restrictive $-18.6$ mag magnitude limit (see Section \ref{sec::data}). We then fit a sinusoid to the median colour values in the form

\begin{equation}
y = A \cdot \cos(f\cdot x) + c,
\label{eq::sinusoidfit}
\end{equation}

where $y$ is the median colour of a given angle bin, $x$ is the central angle of the bin from the major axis of the \ac{BCG}, $A$ is the amplitude of the fit, $f$ is the frequency of the fit and $c$ is the offset which corresponds to the median colour of the bins.

Figures \ref{subfig::1_5R200_radial_aniso_fullpop_low} and \ref{subfig::1_5R200_radial_aniso_low} show the signal for bins of all satellites within $1.5R_{200}$ for our sample using our $R$-band completeness limit. Both panels show the same sinusoidal fit, but Figure \ref{subfig::1_5R200_radial_aniso_fullpop_low} includes the whole population of satellites for visual reference. The red points show the median $(B-R)_{\text{corr.}}$ colour in approximately equal-sized angle bins, with grey dashed (dotted) lines indicating angles along the major (minor) axis. The red shaded region indicates the $1\sigma$ error on $c$ from Equation \ref{eq::sinusoidfit}. From Figure \ref{subfig::1_5R200_radial_aniso_low}, there is a clear anisotropic signal, with peaks in colour along the major axis of the \ac{BCG} and troughs along the minor axis. The fit in Figure \ref{subfig::1_5R200_radial_aniso_low} has a significant amplitude of $A = 0.14\pm0.01$ and a period of $1/f = P = (178.8 \pm 1.6)\degree$. This period is consistent with $180\degree$ demonstrating that our data shows the anisotropic quenching signal first put forward by \citet{Martin-Navarro2021}. Figures \ref{subfig::1_5R200_radial_aniso_fullpop_high} and \ref{subfig::1_5R200_radial_aniso_high} show the same as the top two panels, except this time the satellite galaxies have a more restrictive absolute $R$-band magnitude limit of $-18.6$ mag. The fit for this more conservative sample (blue points and line) has an amplitude of $A = 0.13 \pm 0.01$ and a period of $P = (177.4 \pm 2.2)\degree$, again demonstrating a very significant anisotropic signal that peaks close to the major axis of the \ac{BCG}.

\begin{figure}
        \centering
        \includegraphics[width=\columnwidth, trim=0 0 0 0, clip=true]{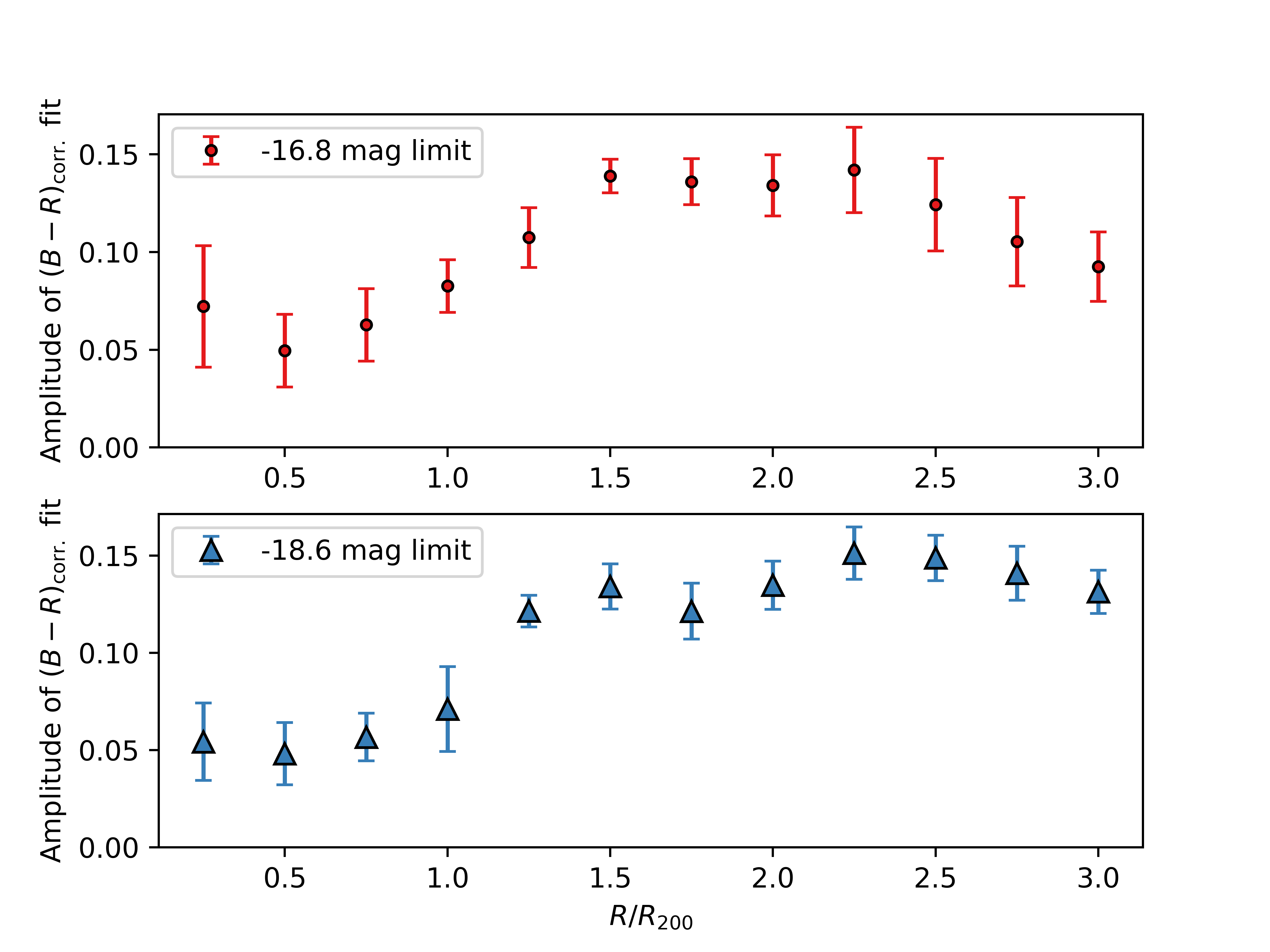}
    \caption[]{The amplitude of the anisotropic quenching signal at different cluster-centric radii. The points represent the amplitude of the fit for all satellites within the corresponding $R/R_{200}$ radius. The errors are the uncertainty on the amplitude of the anisotropic quenching signal. The top panel (red points) indicate the radially evolving amplitude for satellites using our $-16.8$ absolute $R$-band magnitude completeness limit, and the bottom panel (blue triangles) shows this for our $-18.6$ mag limited sample.}
        \label{fig::colouramplitudefits}
\end{figure}

Figure \ref{fig::colourangle_1_5R200} shows the signal for galaxies within a radius of $1.5R_{200}$. However, we also determined the anisotropic signal for all satellites within increasing cluster-centric radii out to $3R_{200}$ at intervals of $0.25R_{200}$. The amplitudes of these fits are shown in Figure \ref{fig::colouramplitudefits}, where each point represents the signal for all galaxies within the corresponding radius. There is no clear radial peak in amplitude for either sample, with the amplitudes for our $-16.8$ mag limited sample remaining approximately constant from $1.5R_{200}$ out to $2.25R_{200}$ before steadily dropping. This drop is likely caused by a contribution of the increasingly isotropic galaxy population at large distances from the centre of the clusters. For the $-18.6$ mag limited sample, there is a sharper rise between $1-1.5R_{200}$ before reaching a plateau at the larger cluster-centric radii of $2.25R_{200}$. As with the $-16.8$ mag limited sample, there is decline in the signal strength at larger radii, though not as strong.

\begin{figure}
        \centering
        \includegraphics[width=\columnwidth, trim=0 0 0 0, clip=true]{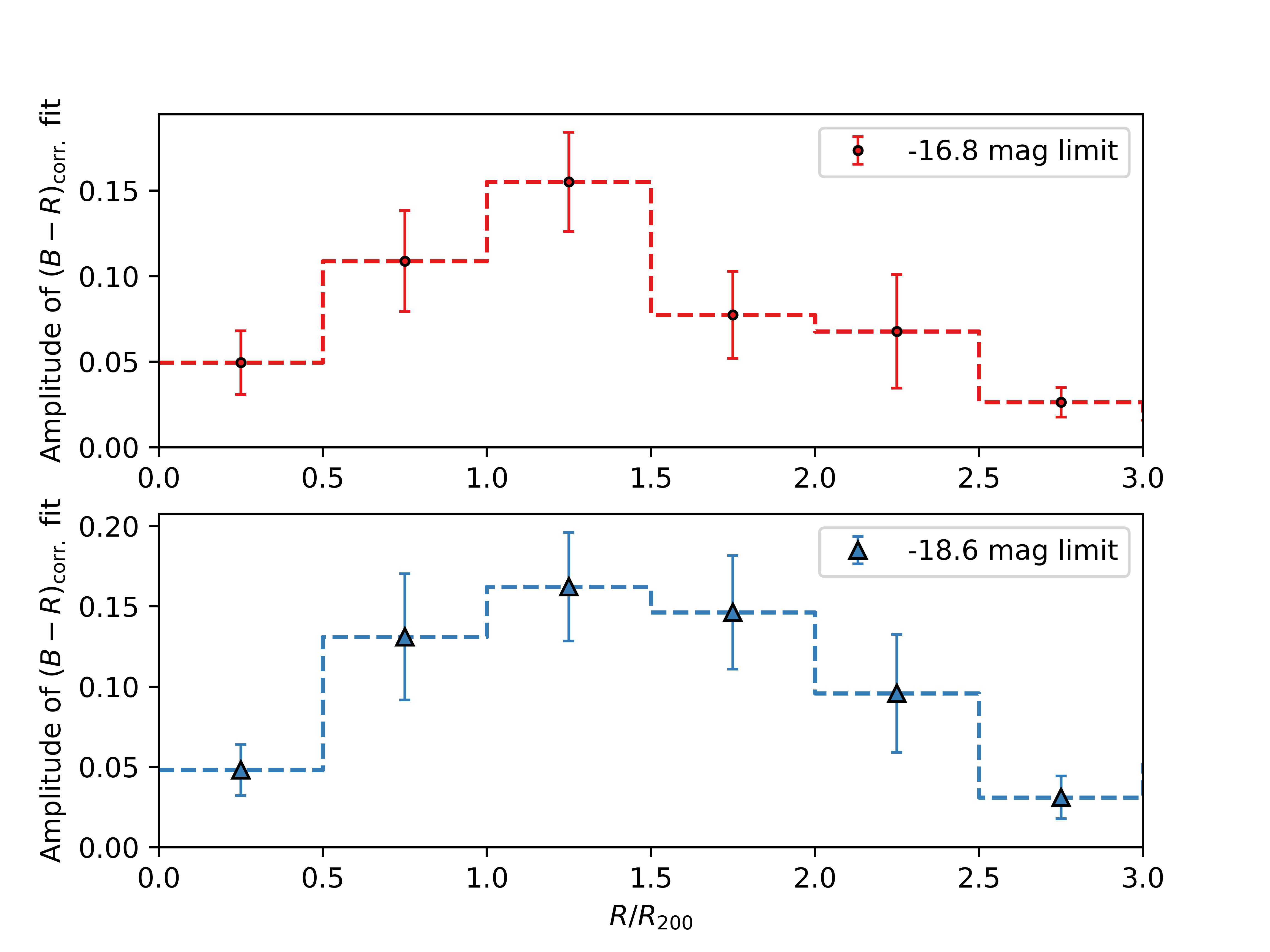}
    \caption[]{The amplitude of the anisotropic quenching signal in colour-angle space within $0.5R_{200}$-wide circular annuli as a function of cluster-centric radius. The points represent the amplitude of the fit within each annulus, and the corresponding $R/R_{200}$ value is the central value of the annulus. The error bars are the uncertainty on the amplitude of the anisotropic quenching signal. The dashed step function represents the range of each circular annulus. The top panel (red points and line) indicates the evolving amplitude for satellites using our $-16.8$ absolute $R$-band magnitude completeness limit, and the bottom panel (blue triangles and line) shows the evolution for our $-18.6$ mag limited sample.}
        \label{fig::colouramplitudefitsannuli}
\end{figure}

To isolate the signal for galaxies at larger cluster-centric radii, and to get a better understanding of the radius at which anisotropic quenching is most prevalent, it is better to analyse the signal within fixed circular annuli going out from the \ac{BCG}. Figure \ref{fig::colouramplitudefitsannuli} shows the amplitude of the sinusoid fit to the anisotropic quenching signal in $0.5R_{200}$ wide annuli. The points represent the amplitude of the signal within each annulus, and the corresponding $R/R_{200}$ represents the central $R_{200}$ value of the bins, which are shown by the dashed lines. From Figure \ref{fig::colouramplitudefitsannuli}, it can be seen that for a completeness limit of $-16.8$ mag in $R$-band there is a significant peak in the amplitude at $1-1.5R_{200}$ of $0.16 \pm 0.03$ before a $\sim2\sigma$ drop to the next annulus. Similarly, for the brighter magnitude limit, the peak is also at $1-1.5R_{200}$ with an amplitude of $0.16 \pm 0.03$, however this is consistent with the amplitude at $1.5-2R_{200}$ of $0.15 \pm 0.04$, clearly indicating a more gradual decrease in signal strength.

\subsection{Passive Galaxy Fraction Relationship}
\label{subsec::anisopassive}

\begin{figure}
        \centering
        \includegraphics[width=\columnwidth, trim=0 0 0 0, clip=true]{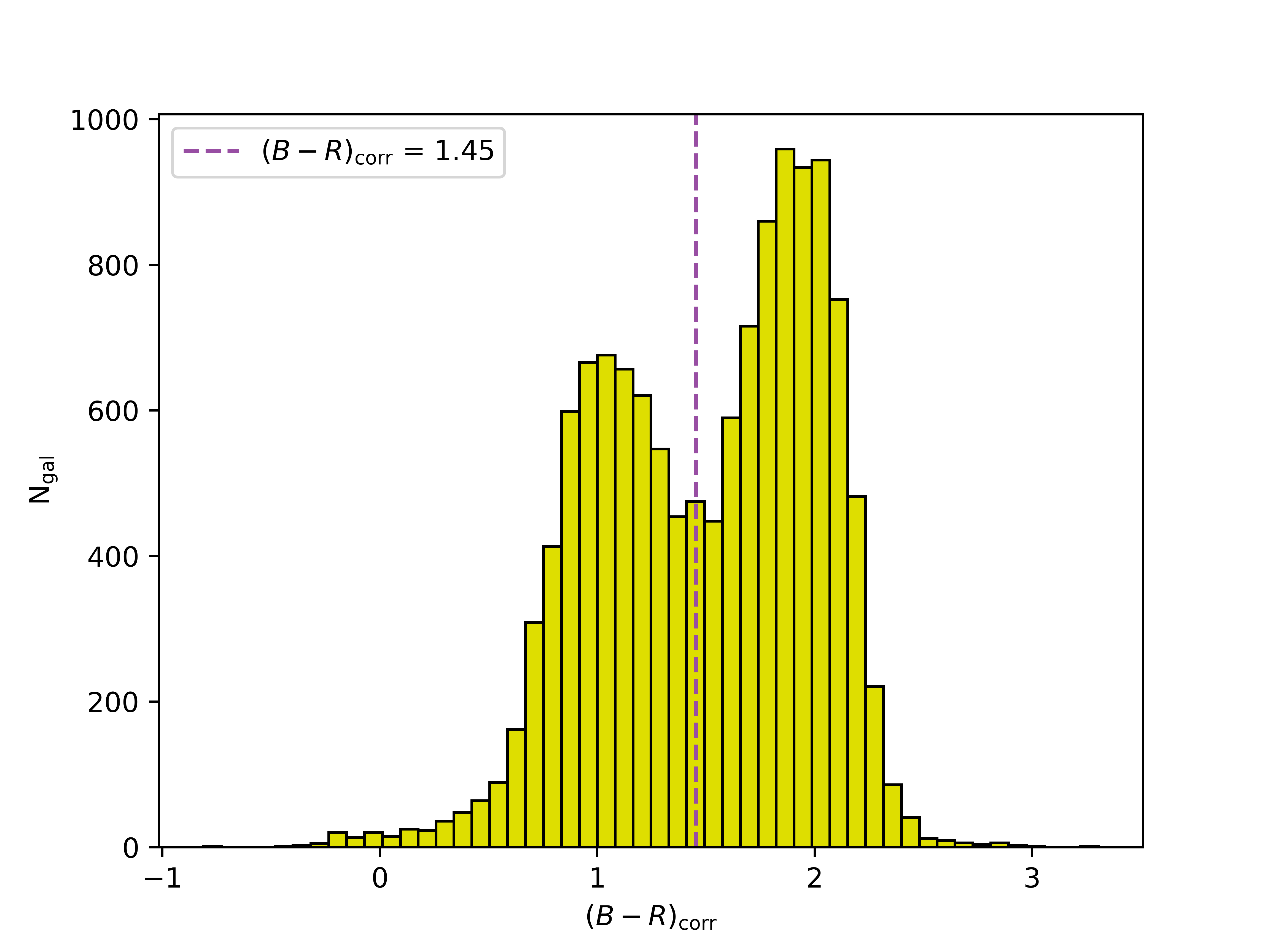}
    \caption[]{The $(B-R)_{\text{corr.}}$ colour distribution of all objects in available CLASH clusters from the Subaru observations of \citet{Umetsu2014} in which $B-R$ is the most applicable colour to distinguish \ac{SF} populations. The purple dashed line indicates the central $(B-R)_{\text{corr.}}$ value between the two distinct peaks in the sample distribution, corresponding to $(B-R)_{\text{corr.}}\approx 1.45$. This value was used to distinguish \ac{SF} satellite galaxies and quiescent ones.}
        \label{fig::colourbinsplit}
\end{figure}

Galaxy quenching describes the process via which a galaxy's star formation is suppressed. Therefore analysis of any possible anisotropic quenching signal would benefit from using the fraction of passive to total galaxies along the major and minor axes. To do this, we split the galaxy populations based on $(B-R)_{\text{corr.}}$ colour for all the available clusters with $B$- and $R$-band data in which $B-R$ is the most applicable colour index for distinguishing \ac{SF} populations given the location of the 4000 \si{\angstrom} break. We then analysed the resulting colour distribution, which included clusters that are not in our final sample for having an unclear \ac{BCG}. Figure \ref{fig::colourbinsplit} shows the distribution of $(B-R)_{\text{corr.}}$ colours for galaxies in the redshift range $z=0.187-0.494$. The purple dashed line represents the middle of the two peaks in the colour distribution and corresponds to $(B-R)_{\text{corr.}} = 1.45$. As a result, we split the satellite galaxies into two populations: \ac{SF} galaxies that have a colour $(B-R)_{\text{corr.}} < 1.45$ and \say{passive} galaxies with $(B-R)_{\text{corr.}} > 1.45$. We then took the same bins as were used in the colour distributions in Figure \ref{fig::colourangle_1_5R200} and divided the number of passive galaxies by the total number of galaxies in each bin to give $f_{\text{pass.}}$.

\begin{figure*}
\centering
\begin{subfigure}{0.45\linewidth}
\includegraphics[width=\linewidth, trim=30 0 30 0]{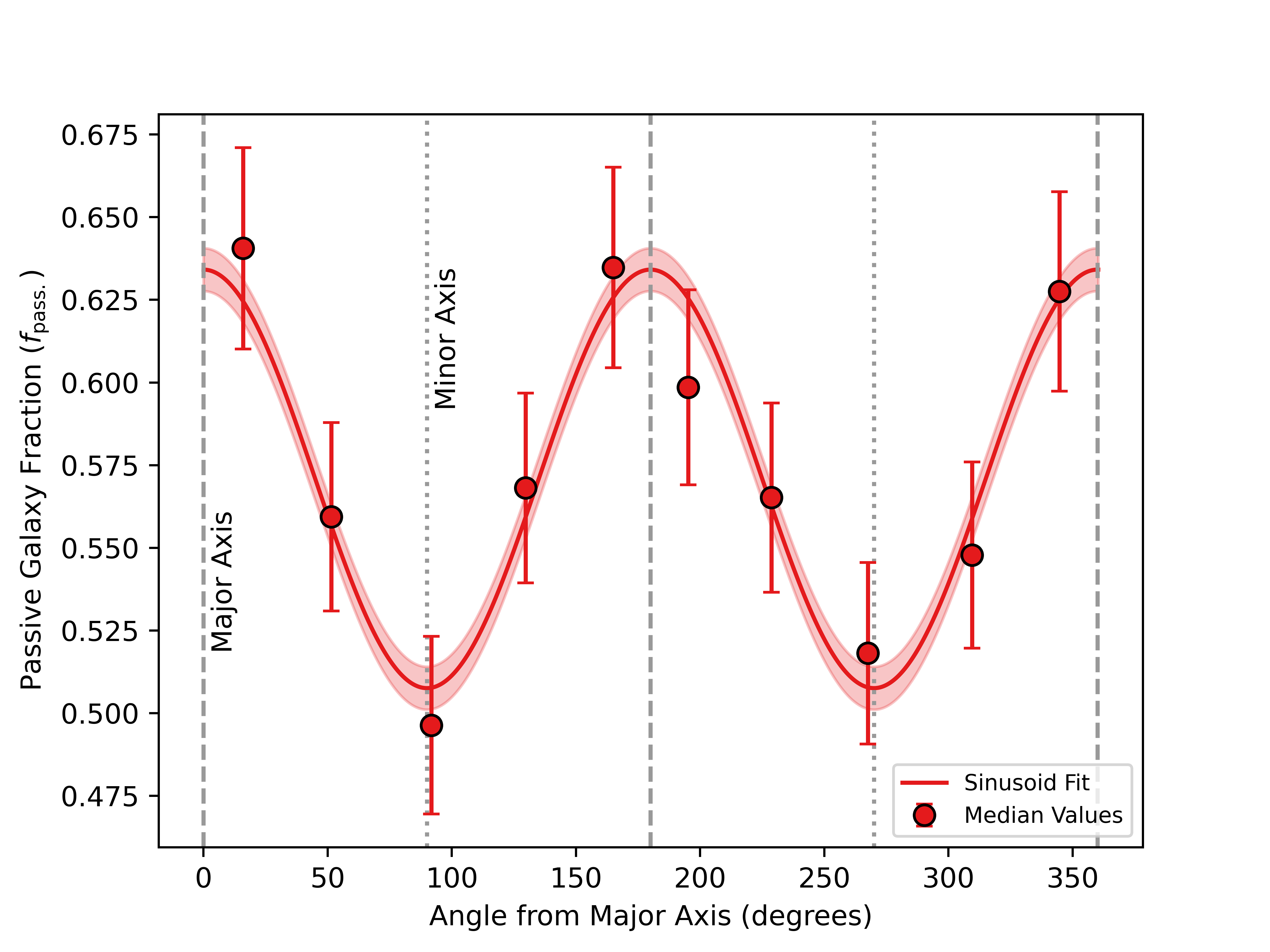}
\caption{ }
\label{subfig::1_5R200_radial_pass_aniso_low}
\end{subfigure}
\hfill
\begin{subfigure}{0.45\linewidth}
\includegraphics[width=\linewidth, trim=30 0 30 0]{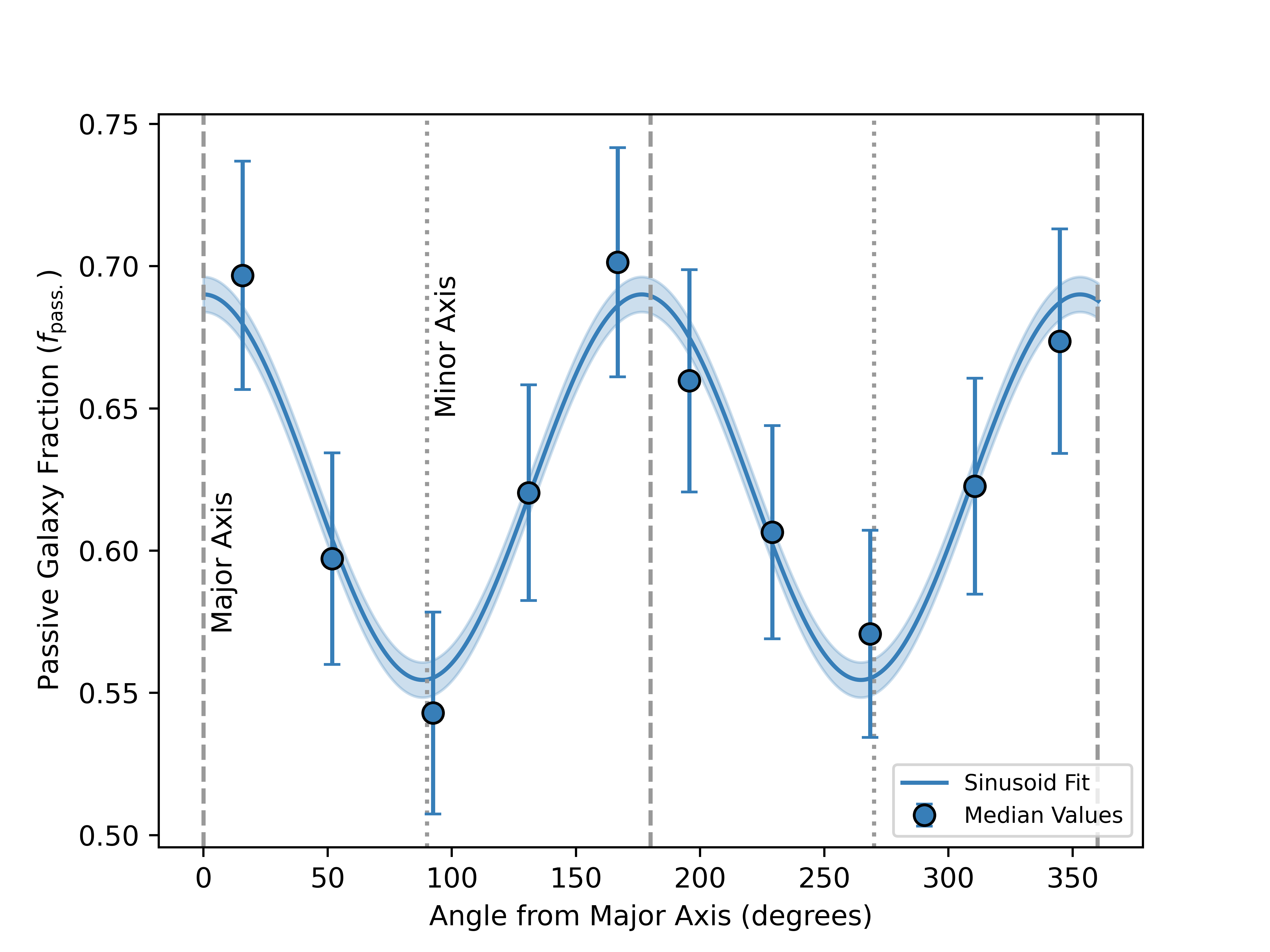}
\caption{ }
\label{subfig::1_5R200_radial_pass_aniso_high}
\end{subfigure}

\caption{Anisotropic quenching signal in $f_{\text{pass.}}$-angle space for satellite galaxies within $1.5R_{200}$ using a $-16.8$ mag completeness limit in $R$-band. The red points and fit in the left panel are the anisotropic signal using an absolute $R$-band completeness limit of $-16.8$ mag, and the blue points and fit in the right panel use a more restrictive limit of $-18.6$ mag. The $f_{\text{pass.}}$ errors for both the red and blue points represent the Poisson error within that bin. The shaded regions represent the $1\sigma$ error of the fits. The grey dashed lines indicate the angle at which satellites are along the major axis of the \ac{BCG}, and the grey dotted lines indicate the angle along the minor axis of the \ac{BCG}.}
\label{fig::passiveangle_1_5R200}
\end{figure*}

Figure \ref{fig::passiveangle_1_5R200} shows $f_{\text{pass.}}$ against angle from the \ac{BCG} major axis for satellite galaxies within $1.5R_{200}$ of the cluster centre. As in Figure \ref{fig::colourangle_1_5R200}, the red (blue) points show the median $(B-R)_{\text{corr.}}$ colour in approximately equal-sized angle bins for our $-16.8$ ($-18.6$) mag $R$-band magnitude limited sample, with the shaded regions indicating the $1\sigma$ error on $c$ from Equation \ref{eq::sinusoidfit}. There is a clear anisotropic signal for $f_{\text{pass.}}$, with an amplitude of $A = 0.063 \pm 0.006$ and period of $P = (180.0\pm 2.6)\degree$ for the $-16.8$ mag sample (average passive galaxy fraction of $\overline{f}_{\text{pass.}} = 0.57\pm0.005$), with $A = 0.068 \pm 0.006$ and period of $P = (176.5\pm 2.4)\degree$ for the $-18.6$ mag sample ($\overline{f}_{\text{pass.}} = 0.62\pm0.005$). Both of these signals are greater than $5\sigma$ and show that the \ac{SFR} of satellites is lower along the major axis of the \ac{BCG}.

\citet{Martin-Navarro2021} also studied the angular dependence of passive galaxy fractions around the \ac{BCG} in $z\sim0.08$ SDSS galaxies. They define \ac{SF} and passive galaxies based on their position on the \ac{SF} main-sequence, fitted in \citet{Martin-Navarro2019} as $\log_{10}(\text{SFR}) = 0.75\log_{10}(M_{*}) - 7.5$. A passive galaxy is defined in their study as one that is offset below the main sequence by more than 1 dex in \ac{SFR}, and \ac{SF} if it is within 1 dex of the main sequence. They found an anisotropic quenching signal in their sample with an amplitude of $0.025\pm0.001$. This is a significant signal, but one that is under half of the strength of those found in our sample of CLASH clusters. They also found an anisotropic quenching signal in IllustrisTNG100 \citep{Weinberger2017,Nelson2019} for SDSS-like galaxies \citep{Rodriguez-Gomez2019} and found the signal had an amplitude of $0.032\pm0.004$ which is in excess of $3\sigma$, but again smaller than for our sample. \citet{Ando2023} analysed the angular dependence of passive galaxy fraction in galaxy clusters using HSC-SSP in redshift bins out to $z\sim1.25$. In their $z=0.25-0.5$ sample - which overlaps best with the redshift sample of our CLASH clusters - they found an anisotropic quenching signal with an amplitude of $0.0167\pm0.0032$ on a sample with an average passive galaxy fraction of $\approx0.82$. They define a galaxy as being quiescent if it has a \ac{sSFR} $< 10^{-11}$ \si{\per\year}. This is a much lower amplitude than the one found in this work, but is still highly significant. These smaller signals from the literature may relate to different properties of the surveys and samples. \citet{Martin-Navarro2021} analysed systems with dark matter halo masses between $\text{M}_{h} = 10^{11.7} - 10^{14.5}$ \si{\solarmass} and \citet{Ando2023} studied clusters with halo masses of $\text{M}_{R_{200}} \approx 10^{13.7} - 10^{14.7}$ \si{\solarmass}, whereas our cluster sample covers a cluster mass range of $\text{M}_{R_{200}} \approx 10^{14.5} - 10^{14.9}$ \si{\solarmass}.

\begin{figure}
        \centering
        \includegraphics[width=\columnwidth, trim=0 0 0 0, clip=true]{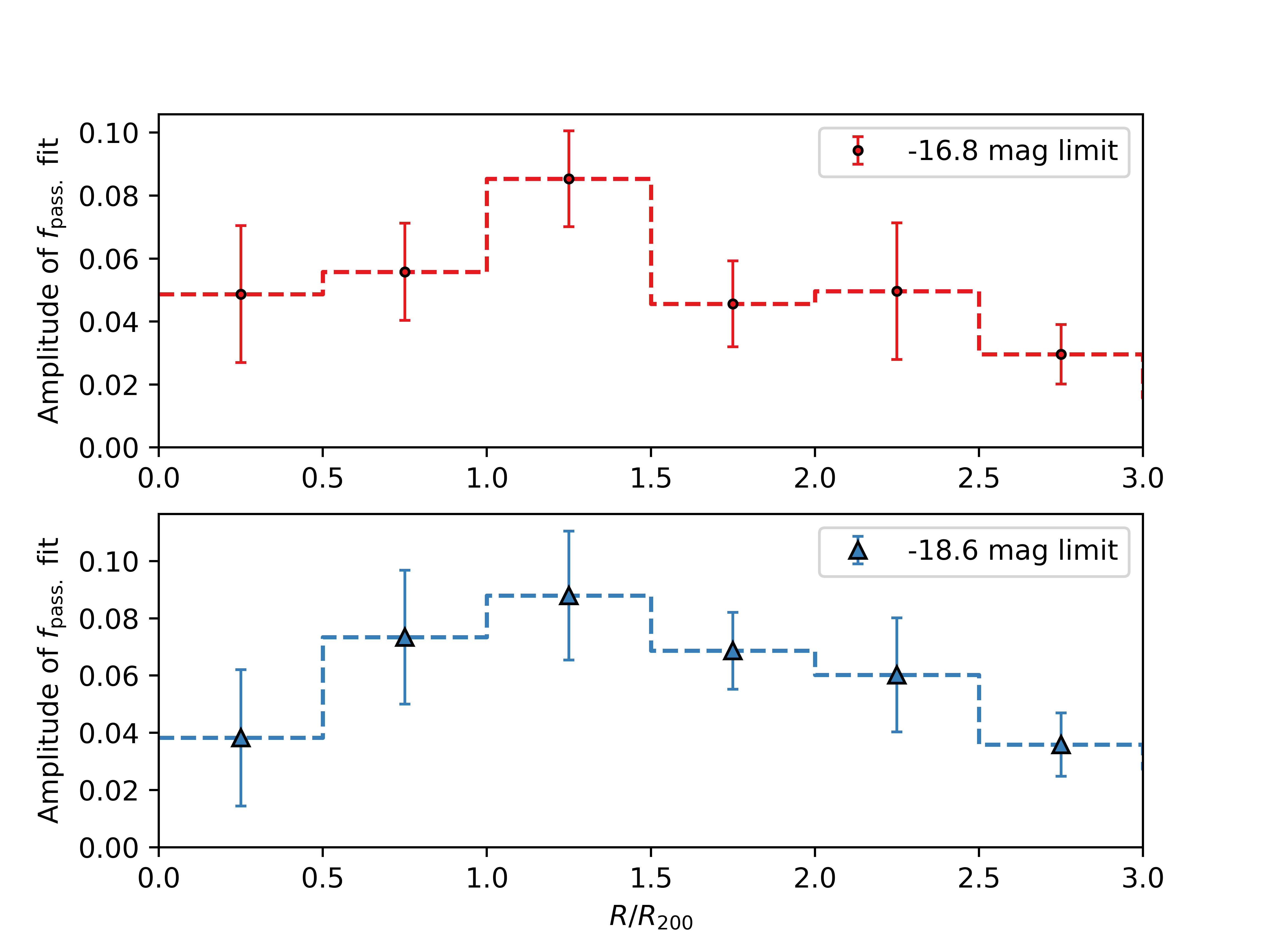}
    \caption[]{The amplitude of the anisotropic quenching signal in $f_{\text{pass.}}$-angle space within $0.5R_{200}$-wide circular annuli as a function of cluster-centric radius. The points represent the amplitude of the fit within each annulus, and the corresponding $R/R_{200}$ value is the central value of the annulus. The errors are the uncertainty on the amplitude of the anisotropic quenching signal. The dashed step function represents the range of each circular annulus. The top panel (red points and line) indicates the radially evolving amplitude for satellites using our $-16.8$ absolute $R$-band magnitude completeness limit, and the bottom panel (blue triangles and line) shows this for our $-18.6$ mag limited sample.}
        \label{fig::passamplitudefitsannuli}
\end{figure}

As in Section \ref{subsec::anisocolour}, we analysed the strength of the anisotropic signal of average $(B-R)_{\text{corr.}}$ in different annular bins to isolate the distance at which the anisotropic quenching signal peaks. Here we look at how $f_{\text{pass.}}$ changes with angle from the \ac{BCG} major axis for both magnitude limits out to $3R_{200}$ at intervals of $0.5R_{200}$-wide annuli. Figure \ref{fig::passamplitudefitsannuli} shows the variation of the amplitude of the signal in these annuli bins. Similarly to Figure \ref{fig::colouramplitudefitsannuli}, the top panel of Figure \ref{fig::passamplitudefitsannuli} shows that, for our $-16.8$ mag sample, there is a peak at $1-1.5R_{200}$ of $0.085\pm0.02$ before dropping at larger radii. The peak for our $-18.6$ mag sample occurs in the same annulus with a peak amplitude of $0.088\pm0.02$, although we note that, as with the bottom panel of Figure \ref{fig::colouramplitudefitsannuli}, the amplitudes between $0.5-2.5R_{200}$ are all in agreement before dropping more significantly at distances $>2.5R_{200}$. This is in contrast to our $-16.8$ mag sample which shows a $\sim2\sigma$ decrease in signal strength directly after the peak.

\subsection{Local Galaxy Number Density}
\label{subsec::numberdensity}

\citet{Ando2023} showed that satellite galaxies are preferentially distributed along the major axis of central cluster galaxies, which would indicate that, along this axis, the number density of satellites will be higher (see also \citealp{Yang2006,Huang2016,Huang2018}). It is also well known that galaxies that are found in denser environments have more suppressed \ac{SFR}s compared to galaxies found in relative isolation (e.g. \citealp{Peng2012,Darvish2016,Crossett2017,Schaefer2019}). As \citet{Stott2022} suggested, it is therefore possible that the difference in quiescent fractions along both axes could be a consequence of the difference in local densities rather than a preferential quenching mechanism acting along the major axis. 

Following a similar method to \citet{Ando2023}, we analyse the local number density of each satellite galaxy across our full sample. We adopt an \emph{n}th-nearest-neighbour method to measure the local surface density of satellite galaxies as it has been shown to be the best estimate of galaxy number densities in massive haloes (see \citealp{Cooper2005} and \citealp{Muldrew2012} for reviews on density measurements). For this work, we define surface density using an average between the 4th- and 5th-nearest-neighbour surface density using

\begin{equation}
 \log_{10}(\Sigma_{\overline{n = 4,5}}) = \frac{1}{2}\log_{10}\left(\frac{4}{\pi d_{n = 4}^{2}}\right) + \frac{1}{2}\log_{10}\left(\frac{5}{\pi d_{n = 5}^{2}}\right),
\label{eq::surfacedensityeq}
\end{equation}
\vspace{0.5mm}

where $d_{n}$ is the distance to the $n$th-nearest-neighbour. We used this averaged surface density approach following other galaxy environment studies (e.g. \citealp{Bamford2009,Ellison2010}) and chose $n=4$ and $n=5$ as they are common values used in the literature for the $n$th-nearest-neighbour method (e.g. \citealp{Etherington2015,Wang2023a,Santucci2023,deVos2024}). Analysis was also conducted using a basic surface density calculation $\Sigma_{n} = n/(\pi d_{n}^{2})$ for $n=4,5$, but the overall conclusions were the same as with using Equation \ref{eq::surfacedensityeq}.

\begin{figure*}
\centering
\begin{subfigure}{0.45\linewidth}
\includegraphics[width=\linewidth, trim=50 0 60 0]{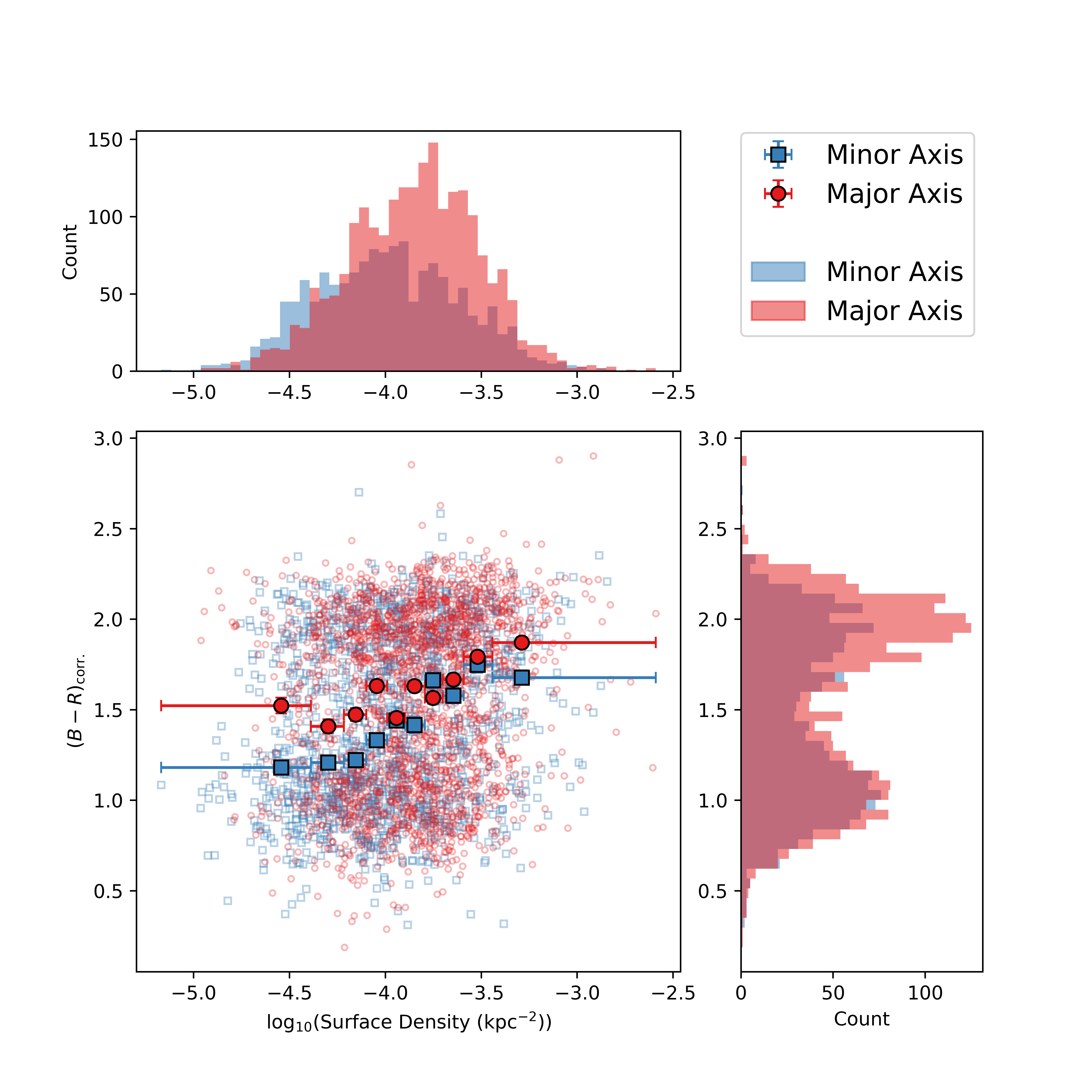}
\caption{ }
\label{subfig::scat_hist_density}
\end{subfigure}
\hfill
\begin{subfigure}{0.45\linewidth}
\includegraphics[width=\linewidth, trim=60 0 60 0]{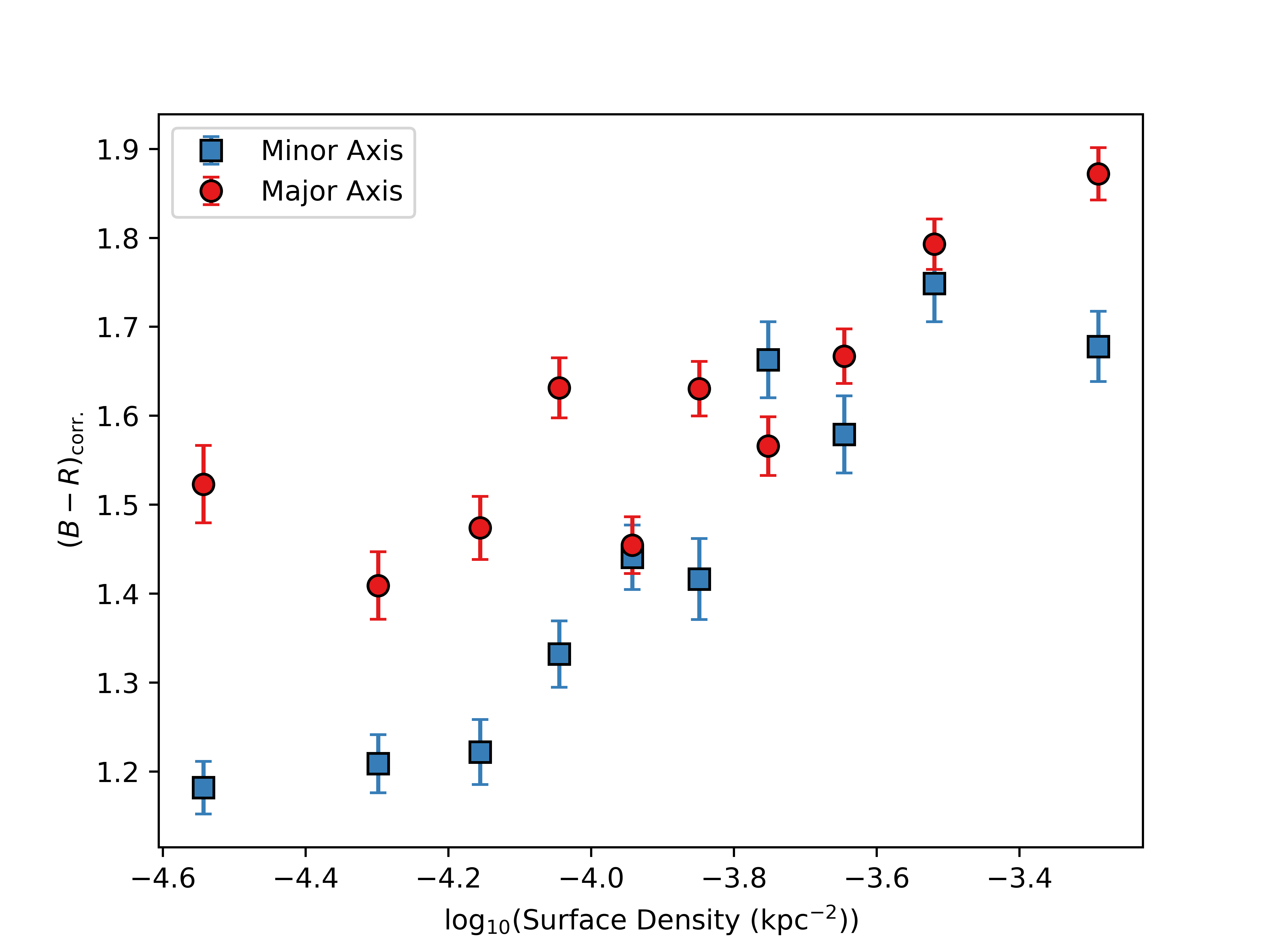}
\caption{ }
\label{subfig::medians_density}
\end{subfigure}

\caption{The distribution of $(B-R)_{\text{corr.}}$ against local surface density of satellite galaxies that are $\pm30\degree$ from either the minor or major axis. In the left panel, the solid red circles and solid blue squares represent the median $(B-R)_{\text{corr.}}$ in surface density bins for galaxies along the major and minor axes respectively. The error on median $(B-R)_{\text{corr.}}$ represents the standard error of the values in each bin, and the error on surface density represents the range of individual values that occupy them. The smaller, empty points represent the individual satellites in this plane. The upper histogram shows the distribution of surface density values, with the red bars indicating satellites along the major axis and the blue bars for those along the minor axis. The right-hand histogram shows the distribution of $(B-R)_{\text{corr.}}$ values along both axes. In the right panel, the points are the same median values as in the left panel, but the individual satellites and surface density errors are removed for additional clarity on the median distribution.}
\label{fig::coloursurfacedensityplot}
\end{figure*}

\begin{figure}
        \centering
        \includegraphics[width=\columnwidth, trim=0 0 0 0, clip=true]{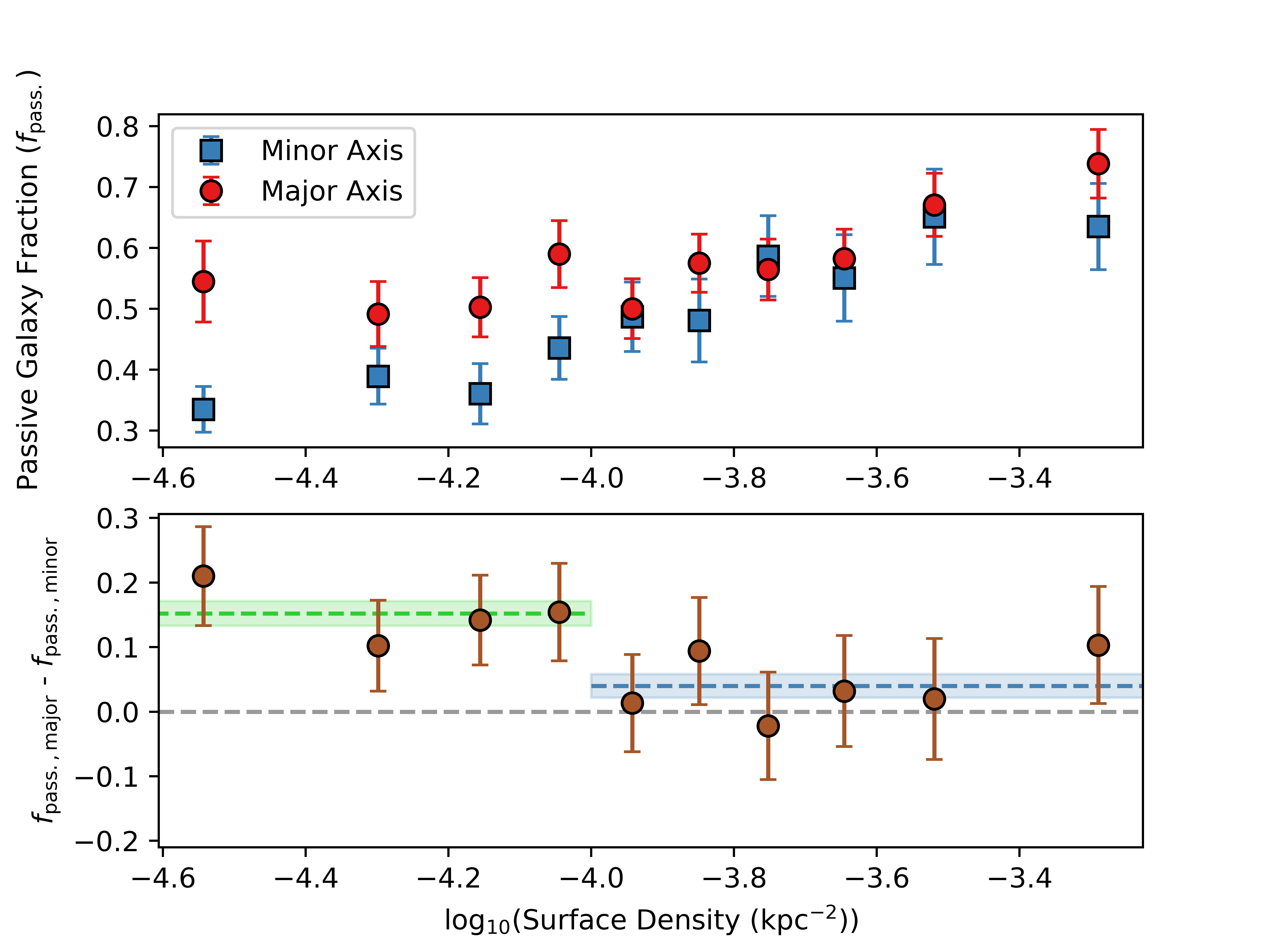}
    \caption[]{The top panel shows the distribution of median $f_{\text{pass.}}$ values in local surface density bins of satellite galaxies that are $\pm30\degree$ from either the minor or major axis. The red points (blue squares) indicate the $f_{\text{pass.}}$ values along the major (minor) axis. The error on $f_{\text{pass.}}$ indicates the Poisson error within each bin. The bottom panel shows the difference between the passive galaxy fraction along the major axis ($f_{\text{pass, major}}$) compared to along the minor axis ($f_{\text{pass, minor}}$) in the same bins. The error is the uncertainty on this difference based on the error of the median $f_{\text{pass.}}$ values in the top panel. The brown dotted line indicates the mean value of $f_{\text{pass, major}} - f_{\text{pass, minor}}$, with the brown shaded region indicating the standard error of the mean. The grey dashed line indicates where there would be no difference between the two axes.}
        \label{fig::passivesurface}
\end{figure}

Figure \ref{fig::coloursurfacedensityplot} shows $(B-R)_{\text{corr.}}$ colour against local surface density calculated from Equation \ref{eq::surfacedensityeq} for cluster galaxies using a $-16.8$ mag completeness limit. We define a galaxy as being along the major or minor axis if it is within $\pm15\degree$ of that axis. We chose this value to be consistent with the analysis of \citet{Ando2023} (though we note that using a range of opening angles from $10\degree$ to $60\degree$ have no effect on our conclusions). From Figure \ref{fig::coloursurfacedensityplot}, we can see that with increasing surface density, the median colour in both major and minor axes broadly increases. However, we can see that for fixed surface density, median $(B-R)_{\text{corr.}}$ is significantly higher along the major axis for surface density bins $<10^{-4.0}$ \si{\per\square\kilo\parsec} before becoming less significant at higher densities. In addition to analysing $(B-R)_{\text{corr.}}$ we also analysed average $f_{\text{pass.}}$ as a function of local surface density in Figure \ref{fig::passivesurface}. This was done by taking the surface density bins of Figure \ref{fig::coloursurfacedensityplot} and calculating $f_{\text{pass.}}$ within each of them for both axes. In the top panel of Figure \ref{fig::passivesurface}, we see an approximately linear increase in $f_{\text{pass.}}$ with surface density, which is a reflection of increased local density suppressing star formation more effectively (e.g. \citealp{Peng2012,Darvish2016,Kawinwanichakij2017}).

Similarly to $(B-R)_{\text{corr.}}$ in Figure \ref{fig::coloursurfacedensityplot}, we see that for fixed local surface densities $<10^{-4.0}$ \si{\per\square\kilo\parsec}, $f_{\text{pass.}}$ is significantly higher along the major axis than the minor axis, with higher densities showing a smaller difference. This is emphasised in the bottom panel of Figure \ref{fig::passivesurface} which shows the difference between $f_{\text{pass.}}$ in the major axis compared to the minor axis. The dotted lines shows the average difference between the densities along the two axes, with the green line showing the average for fixed surface densities $\leq10^{-4.0}$ \si{\per\square\kilo\parsec} of $0.15\pm0.02$, and the blue line showing the difference at $>10^{-4.0}$ \si{\per\square\kilo\parsec} of $0.04\pm0.02$. The difference at low surface densities is significant and suggests that the anisotropic signal we see is caused by some quenching mechanism(s) that preferentially impacts satellites residing along the major axis rather than being a reflection of the different local environments. However, in Section \ref{subsec::preprocessing}, we argue that it is the comparative length of time spent in higher density environments along the two axes that causes the signal. For galaxies that reside in higher surface densities, the much lower difference is likely reflecting the increasingly dominant effects of environmental quenching which occur in high density groups of satellites \citep{Peng2010b}.

\section{Discussion}
\label{sec::discussion}


\subsection{Impact of AGN Outflows}
\label{subsec::radialsignal}


We see that the amplitude of our anisotropic quenching signal using $(B-R)_{\text{corr.}}$ and $f_{\text{pass.}}$ peaks at $\approx1.25R_{200}$ in Figures \ref{fig::colouramplitudefitsannuli} and \ref{fig::passamplitudefitsannuli} respectively, and we analyse the signal out to $3R_{200}$. This is the largest cluster-centric radii that anisotropic quenching has been analysed, and also the first time the peaks of individual sinusoidal amplitudes has been measured within fixed circular annuli extending out from the centre of galaxy clusters. This has allowed us to observe a peak in the signal that other studies that measure the radial dependence of anisotropic quenching have not. For example, \citet{Ando2023} analysed the radial dependence of overall quiescent fractions in annuli along a $30\degree$ wedge from both axes, but here we directly analyse the overall signal in $0.5R_{200}$ annuli from the \ac{BCG}. Other studies have analysed galaxy properties in clusters as a function of cluster-centric radius and linked them to anisotropic quenching, such as green valley fraction and \ac{sSFR} \citep{Jian2023}. From our results in Section \ref{sec::results}, we have shown that anisotropic quenching is not just an effect in the inner regions of galaxy clusters, but instead remains significant - and even rises - out to at least $2.5R_{200}$. The cause of the signal dropping beyond this radius is likely a result of the galaxy population becoming more isotropic, with more distant galaxies (unaffected by environmental effects within the cluster) beginning to smooth the signal.

When they first noted anisotropic quenching, \citet{Martin-Navarro2021} suggested that feedback from the \ac{BCG}'s \ac{AGN} could be responsible. As discussed in Section \ref{sec::intro}, if the radio jets of these powerful \ac{AGN} are aligned with the minor axis of the \ac{BCG}, then they could reduce \ac{ICM} density and make \ac{RPS} less efficient. This follows from X-ray cavities or \say{bubbles} \citep{Boehringer1993,Heckman2014} that are left in the wake of the powerful radio jets. For this to be a plausible explanation for the results we see in this work, these cavities would need to extend out to large cluster-centric radii. However, evidence in the literature suggests the spatial extent of X-ray cavities is relatively small compared to the radii which our anisotropic quenching signal remains significant. In their review on \ac{AGN} feedback, \citet{McNamara2007} used \emph{Chandra} \citep{Weisskopf2002} data from the cluster sample of \citet{Rafferty2006} and found that the detection rates of X-ray cavities in clusters decline rapidly from a peak at 30 \si{\kilo\parsec} from the core of the host galaxy, with very few detected at $>100$ \si{\kilo\parsec}. More recent observations have also confirmed that cavities typically extend to 15-30 \si{\kilo\parsec} from their central host (e.g. \citealp{Hlavacek-Larrondo2013,Birzan2020,Timmerman2022}). Simulations have shown that X-ray cavities can extend to 200-300 \si{\kilo\parsec} from the \ac{BCG} before being suppressed by gas inflows (e.g. \citealp{Cielo2018}), and there are rare examples of cavities extending to $200$ \si{\kilo\parsec} in observations (e.g. \citealp{Nulsen2005,McNamara2005,Wise2007}). The median cluster size of our sample is $R_{200, \text{med.}} = 933 \pm 90$ \si{\kilo\parsec}. With this in mind, taking the higher estimate of 300 \si{\kilo\parsec}, X-ray cavities would only extend to $\approx0.2R_{200}$ at most from the \ac{BCG} of the CLASH clusters. It would require an X-ray cavity to reach $\approx450$ \si{\kilo\parsec} to begin to influence satellites beyond the first annulus of our analysis in Figures \ref{fig::colouramplitudefitsannuli} and \ref{fig::passamplitudefitsannuli}. This does not rule out the influence of \ac{AGN} feedback in the inner-most regions of our clusters, since we do see an anisotropic quenching signal at $R<0.5R_{200}$ and \citet{Martin-Navarro2021} provide evidence from IllustrisTNG100 \citep{Nelson2018} that support their observations which only extend to $\sim0.75R_{200}$ at most \citep{Stott2022}. However, we note that we have also shown that there is no significant difference in the anisotropic quenching signal between the two magnitude limited samples. If the reduced efficiency of \ac{RPS} in \ac{AGN} was the primary driver, then a larger signal would be expected for the $-16.8$ mag sample populated by more lower-mass satellites that are less resistant to the stripping of their gas by the \ac{ICM}.

\ac{AGN}-fuelled X-ray cavities would also need to be common in clusters of galaxies at this redshift, and last long enough to have a noticeable impact on the efficiency of \ac{RPS} of orbiting satellites. \citet{Hlavacek-Larrondo2012} studied \ac{AGN} feedback from the \ac{BCG} in 76 MACS clusters, including three in our sample (MACS1720, MACS1931 and MACS1115). They find clear cavities in only 13 of the 76 clusters, with a further seven that have possible cavities, giving a detection rate of just $\approx26\%$ (though we note that they find cavities in all 3 of the aforementioned clusters which we also analyse here). \citet{Birzan2020} used data from the International Low-Frequency Array (LOFAR; \citealp{VanHaarlem2013}) Two-metre Sky Survey (LoTSS; \citealp{Shimwell2017,Shimwell2019}) to study 42 systems that host possible X-ray cavities, of which 25 are clusters of galaxies. Whilst their study was used to study groups and clusters that already had possible cavities, they found only $\sim54\%$ of their massive clusters show evidence for significant radio mode \ac{AGN} feedback. Other studies have similar cavity detection rates of $\sim30-50\%$ (e.g. \citealp{Panagoulia2014,Shin2016,Olivares2023}). One would expect that, in order to be a primary driver of the anisotropic quenching signal in a population of galaxy clusters, significant X-ray cavities would need to be present in at least a majority of systems. Furthermore, the X-ray cavities need to have existed long enough to influence the properties of the satellites that may interact with them. Using a simple estimate of the average crossing time for our sample of clusters where $R_{200, \text{med.}} = 933$ \si{\kilo\parsec}, and a typical cluster velocity dispersion of $\sigma = 1000$ \si{\kilo\meter\per\second} \citep{Struble1999} then we get an average cluster crossing time of $t_{\text{cross}} \approx \frac{R_{200, \text{med.}}}{\sigma} \approx 0.9$ \si{\giga\year}. However, typical cavity ages are $0.01-0.27$ \si{\giga\year} (e.g. \citealp{Nulsen2005,Hlavacek-Larrondo2012,Hlavacek-Larrondo2013,Macconi2022}). It is therefore unlikely that a significant number of orbiting satellites will have had time to pass through these cavities since they were formed.

There is also direct evidence from simulations that suggest the impact of \ac{AGN} outflows on anisotropic quenching is minimal. \citet{Karp2023} analyse results from the TNG100 cosmological hydrodynamical simulations and UniverseMachine, which is an empirical post-processing algorithm \citep{Behroozi2019}. They run these on the Small MultiDark Planck (SMDPL; \citealp{Rodriguez-Puebla2016}) simulation as part of the MultiDark cosmological dark matter-only suite \citep{Klypin2016} to reassess \citet{Martin-Navarro2021}'s conclusions. They find that \ac{AGN} outflows are not required to explain the anisotropic quenching signal, which is instead fully explained by the hierarchical formation of clusters, which involves the accretion of satellites along cosmic filaments (see Section \ref{subsec::preprocessing}).

\subsection{Large-scale structure and pre-processing}
\label{subsec::preprocessing}

The anisotropic signal remaining significant at cluster-centric radii of $\sim2.5R_{200}$ suggests that the cause of the anisotropic quenching signal could be related to large-scale structure. This is in disagreement with \citet{Ando2023} who found anisotropic quenching is mainly observed within $R_{200}$, although this could be a result of their lower halo mass range or difference in background subtraction. From Figures \ref{fig::coloursurfacedensityplot} and \ref{fig::passivesurface}, it is clear that the anisotropic quenching signal we see in Figures \ref{fig::colourangle_1_5R200} and \ref{fig::passiveangle_1_5R200} respectively are not a reflection of the change in local environment between the major and minor axis. This is because, for fixed surface density values, $(B-R)_{\text{corr.}}$ and $f_{\text{pass.}}$ are significantly higher along the major axis. As such, to probe the possibility of large-scale structure, we analyse the cluster-centric radial evolution of $f_{\text{pass.}}$ directly, and also how local surface density evolves towards the outskirts of the cluster.

\begin{figure}
        \centering
        \includegraphics[width=\columnwidth, trim=0 0 0 0, clip=true]{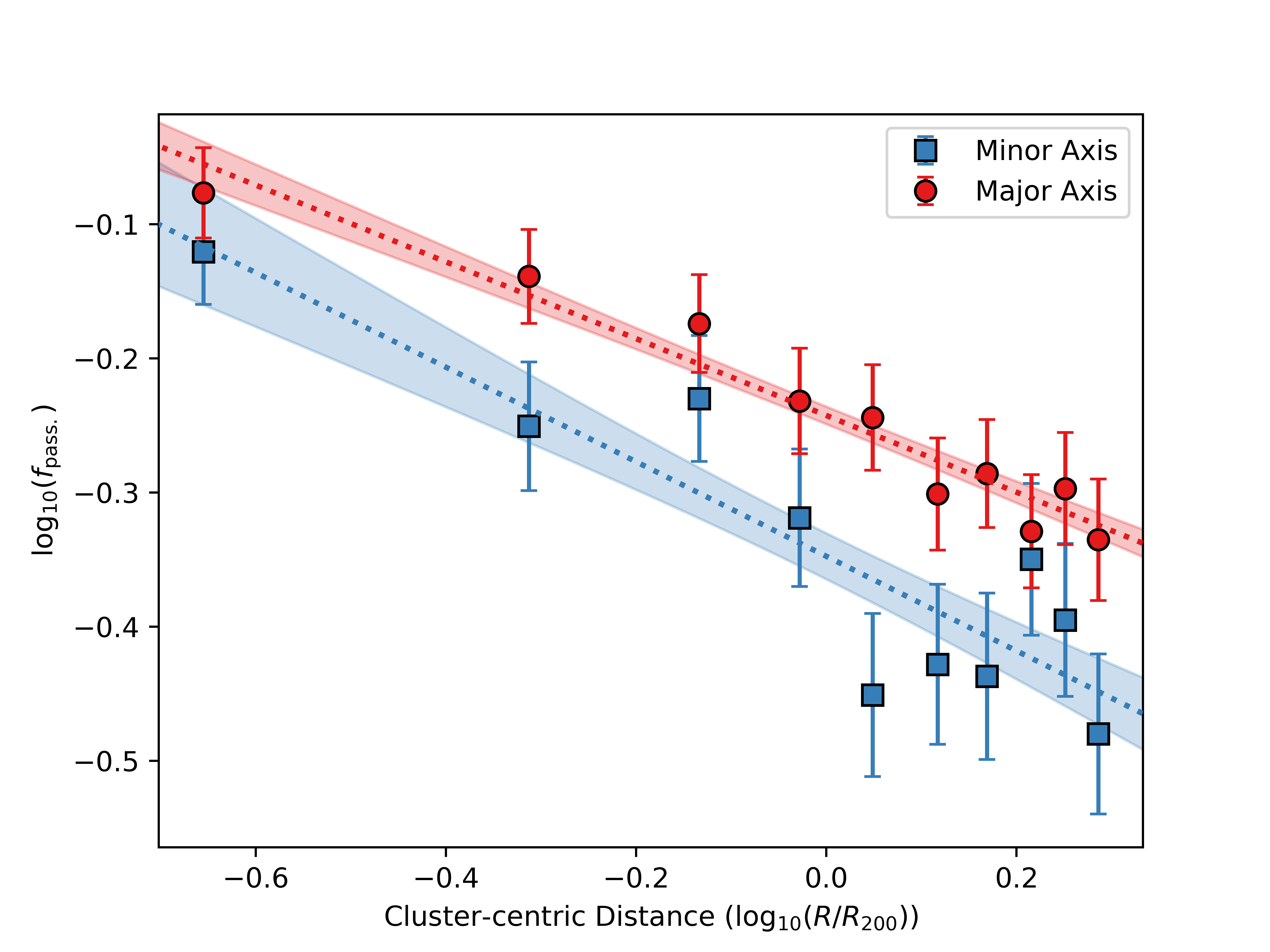}
    \caption[]{The relationship between $\log_{10}(f_{\text{pass.}})$ and distance from the cluster centre in units of $\log_{10}(R/R_{200})$. The red points (blue squares) indicate the median $\log_{10}(f_{\text{pass.}})$ values along the major (minor) axis in cluster-centric distance bins. The error bars represent the Poisson error of the median $f_{\text{pass.}}$ in each bin. The dotted lines represent the best linear fit to the corresponding points, with the shaded region indicating the $1\sigma$ error.}
        \label{fig::distfraglog}
\end{figure}

\begin{figure}
        \centering
        \includegraphics[width=\columnwidth, trim=0 0 0 0, clip=true]{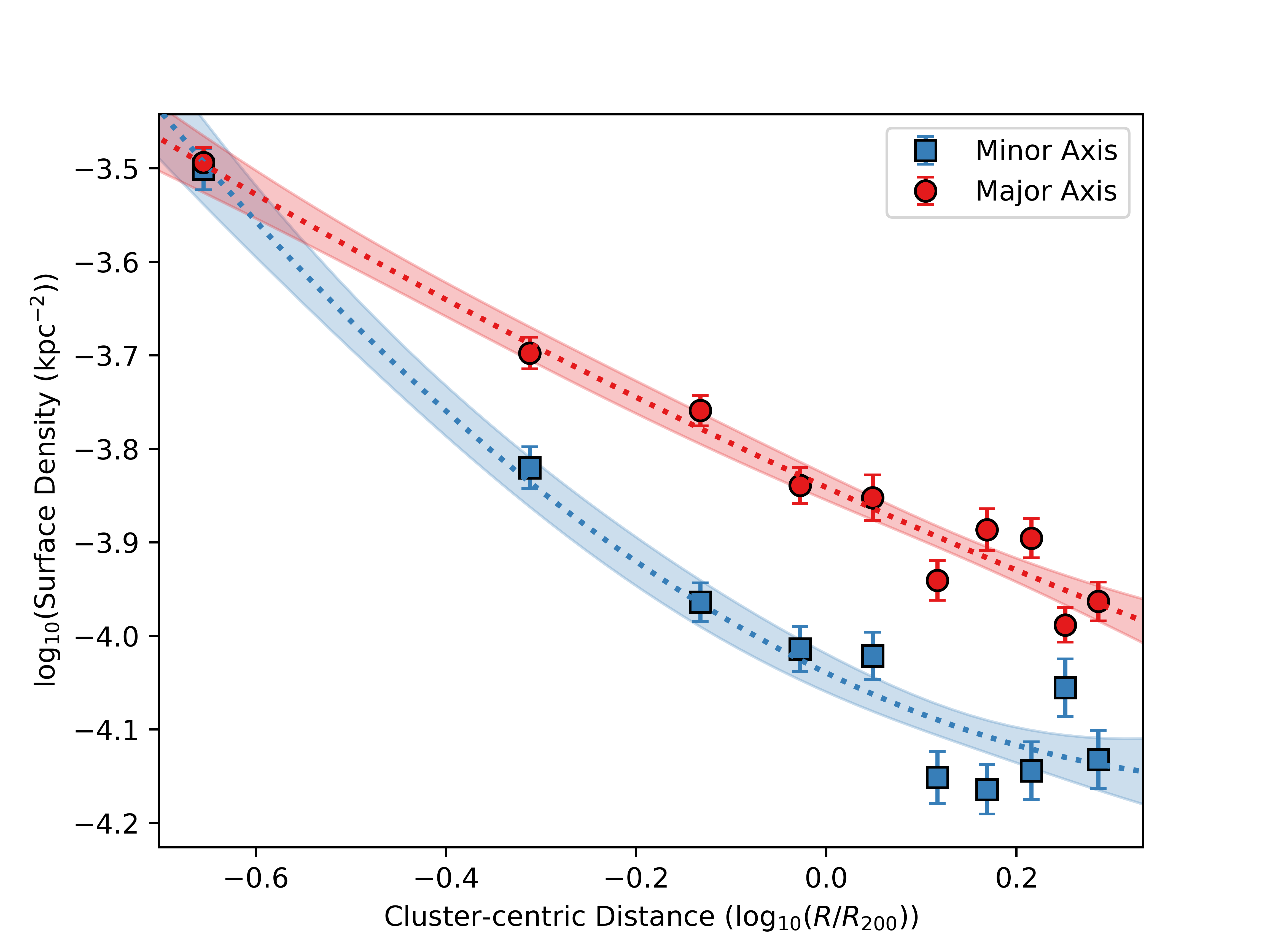}
    \caption[]{The relationship between local surface density and distance from the cluster centre in units of $\log_{10}(R/R_{200})$. The red points (blue squares) indicate the median surface density values along the major (minor) axis in cluster-centric distance bins. The error bars represent the standard error of the median surface density in each bin. The dotted lines represent the best fit in the form $\log_{10}(\Sigma_{\overline{n = 4,5}}) = a \cdot \log_{10}(R/R_{200})^{2} + b \cdot \log_{10}(R/R_{200}) + c$ to the corresponding points. The shaded region indicates the $1\sigma$ error of the fit.}
        \label{fig::densdist}
\end{figure}

Figure \ref{fig::distfraglog} shows the relationship between $f_{\text{pass.}}$ and distance from the centre of the cluster. We can see that $f_{\text{pass.}}$ remains higher along the major axis (red points) compared to the minor axis (blue squares) for fixed distances from the \ac{BCG}. The slope of the minor axis relationship is steeper than that along the major axis, but they agree within errors. The two linear fits (dotted lines) are
\vspace{-0.25mm}
\begin{equation*}
\log_{10}(f_{\text{pass., major}}) = -(0.29\pm0.02) \cdot \log_{10}(R/R_{200}) - (0.24\pm0.01),
\end{equation*}

\begin{equation*}
\log_{10}(f_{\text{pass., minor}}) = -(0.35\pm0.06) \cdot \log_{10}(R/R_{200}) - (0.35\pm0.02).
\end{equation*}

\vspace{0.5mm}\noindent We can see that $f_{\text{pass.}}$ is $\approx0.11$ dex higher along the major axis at $1R_{200}$. A similar result is seen in Figures 5 and 6 of \citet{Ando2023} where they also found passive galaxy fractions are generally higher along the major axis for fixed local overdensity. 

Figure \ref{fig::densdist} shows the relationship between median local surface density and distance from the \ac{BCG}. The fit to both sets of data here is best described by a quadratic in the form
\vspace{-0.25mm}
\begin{equation*}
\log_{10}(\Sigma_{\overline{n = 4,5}}) = a \cdot \log_{10}(R/R_{200})^{2} + b \cdot \log_{10}(R/R_{200}) + c,
\end{equation*}

\vspace{0.5mm}\noindent though we note that the best fit to the major axis data points was also well described by a linear fit. From Figure \ref{fig::densdist}, we see a decline in the local surface density along both axes out to $2R_{200}$. However, we see that the surface densities fall more rapidly along the minor axis before flattening at $\approx1.3R_{200}$, though they are significantly below the equivalent surface densities along the major axis. This shows the median density to be generally much higher along the major axis, demonstrating that the results seen in Figures \ref{fig::coloursurfacedensityplot} and \ref{fig::passivesurface} are mostly driven by the fact galaxies along the major axis have consistently resided in higher density regions compared to the corresponding galaxies along the minor axis. Galaxies infalling along the major axis have therefore spent more time in high density regions and have had greater opportunity to be pre-processed as a result.

In a hierarchical framework, pre-processing describes the scenario in which satellite galaxies that have previously spent time as part of a group of galaxies before their infall into a larger cluster experience a degree of environmental quenching before interacting with the dense \ac{ICM} \citep{Zabludoff1998,Fujita2004}. Smaller groups, or individual galaxies, may also have their star formation suppressed as they travel along cosmic filaments where other quenching mechanisms may be enhanced compared to the field (e.g. \citealp{Sarron2019,Kraljic2020,Donnan2022,Hoosain2024}).

There is strong observational evidence for the pre-processing of satellite galaxies in clusters. \citet{Cortese2004} discovered an infalling group belonging to Abell1367 (see also \citealp{Sakai2002} and \citealp{Iglesias-Paramo2002}), and in a follow-up analysis, \citet{Cortese2006} were able to reconstruct the evolutionary history of the group. They discovered that in the process of infalling into Abell1367, tidal interactions within the group resulted in gas stripping and a weakened potential well that made \ac{RPS} by the \ac{ICM} more efficient on the outskirts of the cluster. \citet{Olave-Rojas2018} studied two $z\sim0.4$ CLASH clusters - MACS0416 and MACS1206, the latter of which is part of our cluster sample - to determine the environmental effects in substructures of galaxies in and around them. In the substructures they identified, they found that the fraction of red galaxies (a good proxy for our $f_{\text{pass.}}$ parameter) is higher in substructures on the outskirts of the clusters than the field, which they conclude is a clear sign of pre-processing in action. They go further and find that the quenching efficiency in groups at cluster-centric radii of $R \geq R_{200}$ is similar to that of the main cluster itself. This supports the idea that pre-processing can lead to already-red galaxies entering larger structures. \citet{Estrada2023} support the results of \citet{Olave-Rojas2018} by finding three substructures at $\sim5R_{200}$ from the centre of MACS0416 which have galaxy populations of similar density, luminosity and colour as those in the core of the cluster, suggesting they are already evolved prior to their incorporation. More observational evidence of pre-processing on the outskirts of clusters ($R \gtrsim R_{200}$), and other high mass haloes, exists in the literature (e.g. \citealp{Gill2005,McGee2009,Dressler2013,Bianconi2018,Dzudzar2019,Einasto2020,Werner2021}), as well evidence in simulations (e.g. \citealp{Yi2013,Joshi2017,Han2018,Bakels2021}).

Massive galaxy clusters are found at the nodes of cosmic filaments \citep{Bond1996} which feed infalling, often pre-processed, galaxies into the cluster (as discussed above; e.g. \citealp{Ebeling2004,Martinez2016,Salerno2019}). As demonstrated in Figure \ref{fig::densdist}, there is tentative evidence that the major axis of the \ac{BCG} is preferentially aligned with the large-scale structure. This was first noted by \citet{Argyres1986}, and further confirmed by \citet{Lambas1988}, who found a correlation between the \ac{BCG} major axis and Lick galaxy counts \citep{Seldner1977} in Abell clusters. Recently, \citet{Smith2023} found that the \ac{BCG}s in their sample of 211 X-ray selected clusters are significantly preferentially aligned with the large-scale structure out to as far as $10R_{200}$ from the cluster centre. In a similar result, \citet{Paz2011} found that there was a correlation between the overall shapes of galaxy groups and their surrounding galaxies out to distances of $\approx30$ \si{\mega\parsec}. Additionally, some studies have gone further and found that, as well as aligning with cosmic filaments, clusters can also align with other nearby clusters and thereby the filaments that may connect them (see \citealp{Govoni2019}). \citet{VanUitert2017} found that redMaPPer clusters tend to align with neighbouring clusters, suggesting the systems share what they dub a \say{common alignment mechanism} (see also \citealp{Smargon2012}).

Additionally, some studies have found that there exists a significant population of backsplash galaxies that occupy the outer regions of relaxed clusters (\citealp{Gill2005}; see also \citealp{Balogh2000a,Mamon2004,Knebe2011}). For example, \citet{Kuchner2021} found that as many $40-60\%$ of satellites at $1-1.5R_{200}$ of relaxed clusters are backsplash galaxies in {\sc The ThreeHundred}\footnote{https://www.nottingham.ac.uk/astronomy/The300/index.php} cosmological zoom-in simulations \citep{Cui2018}. We suggest that the peak anisotropic quenching signal found at $1-1.5R_{200}$ may be due to a build-up of backsplash galaxies along the major axis.


Our results show that galaxies on the major axis have been in high density regions for longer and therefore have experienced more pre-processing. This reconciles all previous works on anisotropic quenching in massive clusters, as our results suggest it is simply pre-processing in major filaments aligned with the BCG.

\section{Conclusions}
\label{sec::conclusions}

In this work, we have investigated the observed phenomenon dubbed anisotropic quenching (also known as `angular conformity' or `angular segregation') of satellite galaxies in 11 CLASH clusters at $z\sim0.36$. We analysed galaxy colour $(B-R)_{\text{corr.}}$ and passive galaxy fractions ($f_{\text{pass.}}$) as a function of their orientation angle from the major axis of the \ac{BCG}. This analysis was done on two samples of galaxies determined by setting different $R$-band magnitude limits to see if the effects of anisotropic quenching differ between galaxy populations. The \ac{SFR} of a satellite galaxy is highly dependent on its local environment \citep{Peng2010b}, therefore we also measured $(B-R)_{\text{corr.}}$ and $f_{\text{pass.}}$ as a function of local surface density using an $n$th-nearest neighbour method (see Equation \ref{eq::surfacedensityeq}) along both axes to determine if the signal was caused by a difference in environment between them. Our results are summarised as follows:

\begin{enumerate}[label=\roman*), leftmargin=*]

    \item From Figure \ref{fig::colourangle_1_5R200}, we find there is an anisotropic angular distribution of $(B-R)_{\text{corr.}}$ from the \ac{BCG} major axis in the CLASH clusters. Using a $-16.8$ ($-18.6$) mag $R$-band magnitude limit, we find the signal is well described by cosine fit (see Equation \ref{eq::sinusoidfit}), where the amplitude of the signal is $A = 0.14\pm0.01$ ($A = 0.13\pm0.01$) and peaks along the major axis. This is a significant amplitude and is among the largest found when analysing the angular evolution of satellite colour from the \ac{BCG} (e.g. \citealp{Stott2022}). The period of the signal is $P = 178.8\degree \pm 1.6\degree$ ($P = 177.4\degree \pm 2.2\degree$).

    \item To analyse anisotropic quenching more directly, it is more appropriate to analyse differences in $f_{\text{pass.}}$ and, from Figure \ref{fig::passiveangle_1_5R200}, we find a clear anisotropic signal. Using Equation \ref{eq::sinusoidfit} to fit the signal to our $-16.8$ ($-18.6$) mag magnitude limited sample, we find an amplitude of $A = 0.063\pm0.006$ ($A = 0.068\pm0.006$) with a period of $P = 180.0\degree \pm 2.6\degree$ ($P = 176.5\degree \pm 2.4\degree$). The amplitude of this signal is in excess of $5\sigma$ with a fitted period consistent with $\approx180\degree$ which peaks along the major axis. This is a clear sign that there is anisotropic quenching in our sample of galaxy clusters.

    \item We find little difference in the strength of the anisotropic quenching signal between our two magnitude limited samples. This indicates that the driver of the signal may not be ram pressure stripping, as we would expect to see a larger impact on the more complete sample.

    \item We measured the amplitude of the anisotropic quenching signal in $0.5R_{200}$-diameter circular annuli going out from the centre of the cluster up to $3R_{200}$. This is the largest cluster-centric radius in which anisotropic quenching has been observed, and also marks the first time the amplitude of the sinusoidal signal in individual annuli has been measured. We find that the amplitude of the $f_{\text{pass.}}$ fit is significant out to at least $2.5R_{200}$ where the amplitude begins to drop. The peak of the fit is $\sim1.25R_{200}$ for both magnitude limited samples, though there is a much sharper drop for our $-16.8$ mag limited sample at $>1.5R_{200}$.
    
    \item We find that $f_{\text{pass.}}$ increases with increasing local surface density, in line with expectations. However, we find that $f_{\text{pass.}}$ is $0.15\pm0.02$ higher along the major axis of the \ac{BCG} for surface density values $\lesssim10^{-4.0}$ \si{\per\square\kilo\parsec}. This is a significant difference which suggests that differences in local density between the major and minor axes are not the primary driver of anisotropic quenching. For surface densities $\gtrsim10^{-4.0}$ \si{\per\square\kilo\parsec}, the difference in $f_{\text{pass.}}$ drops to $0.04\pm0.02$, suggesting the effects of environmental quenching in these denser groups begins to dominate.

    \item We analysed how surface densities evolve from the centre of the cluster. We find that both $f_{\text{pass.}}$ (see Figure \ref{fig::distfraglog}) and local surface densities (see Figure \ref{fig::densdist}) are higher along the major axis than the minor axis. Both parameters drop more quickly along the minor axis than the major axis. We interpret this as galaxies, which are being preferentially fed along the major axis, infall into the cluster in groups that have been pre-processed over a longer time. This reconciles previous works on anisotropic quenching in massive clusters.

\end{enumerate}

A recent paper from \citet{Zakharova2025} came out during the final review stage of this study which investigated anisotropic quenching in satellite galaxies of $\text{M}_{h} = 10^{12} - 10^{14.2}$ \si{\solarmass} haloes in the IllustrisTNG simulations, specifically the TNG100-1 magneto-hydrodynamical model. They analyse the anisotropic quenching signal out to $5R_{200}$, separating between young and old populations. They find that satellite galaxies preferentially infall into the haloes along the major axis of the central galaxy through cosmic filaments that themselves tend to align with the major axis. This agrees with our interpretation here based on observations of the CLASH clusters. They conclude that the anisotropic quenching signal emerges at the point young satellites infall into the larger halo and that this signal, similar to our work, is significant out to large radii from the centre of the halo ($5R_{200}$). Their findings further agree with our conclusions by showing that \ac{AGN} feedback cannot explain the signal at these large radii. Whilst based on an analysis with a lower average halo mass than our sample, we believe the remarkable agreement between our conclusions and those of \citet{Zakharova2025} further strengthen the interpretation that anisotropic quenching is a result of large scale structure.

A logical extension to the anisotropic quenching signal we observe is that galaxies along the major axis may be more elliptical, and those along the minor axis may be more disc-like. Existing observations of the satellites in our sample could confirm if this is the case here, and simulations can be used to test if this is a regular occurrence. If it is, then this may have implications for weak gravitational lensing measurements, with contaminants from the cluster having a different shape depending on their orientation about the BCG (see \citealp{Kirk2015} for a review on weak lensing). Additionally, as discussed in Section \ref{subsec::preprocessing}, there have been simulations that support the idea of pre-processing of infalling cluster galaxies, as well as those that show that the overall shape of clusters align with the cosmic filaments that feed them these pre-processed groups. In the future, to probe the ideas presented in this paper, further simulations of infalling galaxy groups should be explored with an additional focus on whether pre-processed galaxies remain in orbit along the major axis long enough for an anisotropic quenching signal to be readily apparent as a result of their incorporation into the cluster. Future studies should also focus on the impact of backsplash galaxies on this signal.

\section*{Acknowledgements}

{The authors would like to thank the anonymous referee for their constructive comments which has made the analysis of this work more robust and improved the paper.

This work makes use of {\sc AstroPy}\footnote{http://www.astropy.org}, a community-developed core Python package for Astronomy \citep{TheAstropyCollaboration2013,TheAstropyCollaboration2018}, as well as the {\sc NumPy} \citep{Harris2020} and {\sc SciPy} \citep{Virtanen2020} packages (see also \citealp{Oliphant2007}). All plots were created using the {\sc matplotlib} 2D graphics Python package \citep{Hunter2007}.

HMOS gratefully acknowledges support from an STFC PhD studentship and the Faculty of Science and Technology at Lancaster University.

For the purpose of open access, the authors have applied a Creative Commons attribution (CC BY) licence to any author-accepted manuscript version arising.}

\section*{Data availability}

The data underlying this article are publicly available from the CLASH website (\url{https://archive.stsci.edu/prepds/clash/}) as well as the Mikulski Archive for Space Telescopes (MAST, \url{https://archive.stsci.edu/missions-and-data/hst}).




\bibliographystyle{mnras}
\bibliography{MNRAS_HMO_Aniso_Quench.bib} 






\bsp    
\label{lastpage}
\end{document}